\documentclass[a4,11pt]{article}
\usepackage{graphicx} 
\usepackage{float}
\usepackage{amsmath,amsthm}
\usepackage{amsfonts}
\usepackage{latexsym}
\usepackage{amssymb}
\usepackage{setspace}
\usepackage{comment}
\usepackage{ascmac}
\usepackage{cases}
\usepackage{color}
\makeatletter
 
  \@addtoreset{equation}{section}
 \makeatother

\begin{document}
\title{{Equilibrium Price Formation with a Major Player\\ 
and its Mean Field Limit}~\footnote{
Forthcoming in {\it ESAIM: Control, Optimization and Calculus of Variations}.
All the contents expressed in this research are solely those of the author and do not represent any views or 
opinions of any institutions. The author is not responsible or liable in any manner for any losses and/or damages caused by the use of any contents in this research.
}}

\author{Masaaki Fujii\footnote{Quantitative Finance Course, Graduate School of Economics, The University of Tokyo. }, \quad
Akihiko Takahashi\footnote{Quantitative Finance Course, Graduate School of Economics, The University of Tokyo. }
}
\date{ 
First version:  22 February, 2021\\
This version: 14 February,  2022
}
\maketitle



\newtheorem{definition}{Definition}[section]
\newtheorem{assumption}{Assumption}[section]
\newtheorem{condition}{$[$ C}
\newtheorem{lemma}{Lemma}[section]
\newtheorem{proposition}{Proposition}[section]
\newtheorem{theorem}{Theorem}[section]
\newtheorem{remark}{Remark}[section]
\newtheorem{example}{Example}[section]
\newtheorem{corollary}{Corollary}[section]
\def\n{{\bf n}}
\def\A{{\bf A}}
\def\B{{\bf B}}
\def\C{{\bf C}}
\def\D{{\bf D}}
\def\E{{\bf E}}
\def\F{{\bf F}}
\def\G{{\bf G}}
\def\H{{\bf H}}
\def\I{{\bf I}}
\def\J{{\bf J}}
\def\K{{\bf K}}
\def\L{{\bf L}}
\def\M{{\bf M}}
\def\N{{\bf N}}
\def\O{{\bf O}}
\def\P{{\bf P}}
\def\Q{{\bf Q}}
\def\R{{\bf R}}
\def\S{{\bf S}}
\def\T{{\bf T}}
\def\U{{\bf U}}
\def\V{{\bf V}}
\def\W{{\bf W}}
\def\X{{\bf X}}
\def\Y{{\bf Y}}
\def\Z{{\bf Z}}
\def\cala{{\cal A}}
\def\calb{{\cal B}}
\def\calc{{\cal C}}
\def\cald{{\cal D}}
\def\cale{{\cal E}}
\def\calf{{\cal F}}
\def\calg{{\cal G}}
\def\calh{{\cal H}}
\def\cali{{\cal I}}
\def\calj{{\cal J}}
\def\calk{{\cal K}}
\def\call{{\cal L}}
\def\calm{{\cal M}}
\def\caln{{\cal N}}
\def\calo{{\cal O}}
\def\calp{{\cal P}}
\def\calq{{\cal Q}}
\def\calr{{\cal R}}
\def\cals{{\cal S}}
\def\calt{{\cal T}}
\def\calu{{\cal U}}
\def\calv{{\cal V}}
\def\calw{{\cal W}}
\def\calx{{\cal X}}
\def\caly{{\cal Y}}
\def\calz{{\cal Z}}
%
\def\sskip{\hspace{0.5cm}}
\def\simleq{ \raisebox{-.7ex}{\em $\stackrel{{\textstyle <}}{\sim}$} }
\def\leqsim{ \raisebox{-.7ex}{\em $\stackrel{{\textstyle <}}{\sim}$} }
\def\ep{\epsilon}
\def\half{\frac{1}{2}}
\def\iku{\rightarrow}
\def\Iku{\Rightarrow}
\def\ikup{\rightarrow^{p}}
\def\inclusion{\hookrightarrow}
\def\cadlag{c\`adl\`ag\ }
\def\up{\uparrow}
\def\down{\downarrow}
\def\doti{\Leftrightarrow}
\def\douti{\Leftrightarrow}
\def\dochi{\Leftrightarrow}
\def\douchi{\Leftrightarrow}%
\def\yy{\\ && \nonumber \\}
\def\y{\vspace*{3mm}\\}
\def\nn{\nonumber}
\def\be{\begin{equation}}
\def\ee{\end{equation}}
\def\bea{\begin{eqnarray}}
\def\eea{\end{eqnarray}}
\def\beas{\begin{eqnarray*}}
\def\eeas{\end{eqnarray*}}
%
\def\hd{\hat{D}}
\def\hv{\hat{V}}
\def\hsd{{\hat{d}}}
\def\hx{\hat{X}}
\def\hsx{\hat{x}}
\def\bsx{\bar{x}}
\def\bsd{{\bar{d}}}
\def\bx{\bar{X}}
\def\ba{\bar{A}}
\def\bb{\bar{B}}
\def\bc{\bar{C}}
\def\bv{\bar{V}}
\def\balpha{\bar{\alpha}}
\def\bbalpha{\bar{\bar{\alpha}}}
\def\combi{\l(\begin{array}{c}\alpha\\ \beta \end{array}\r)}
\def\f{^{(1)}}
\def\s{^{(2)}}
\def\ss{^{(2)*}}
\def\l{\left}
\def\r{\right}
\def\a{\alpha}
\def\b{\beta}
\def\L{\Lambda}

\def\calf{{\cal F}}
\def\wt{\widetilde}
\def\mbb{\mathbb}
\def\ol{\overline}
\def\ul{\underline}
\def\sign{{\rm{sign}}}
\def\wh{\widehat}
\def\mg{\mathfrak}
\def\display{\displaystyle}

\def\vr{\varrho}
\def\p{\prime}
\def\pp{^\prime}
\def\ep{\epsilon}
\def\vep{\varepsilon}
\def\del{\delta}
\def\Del{\Delta}
\def\red{\textcolor{red}}
\def\Ns{^{(N)}}
\def\ch{\check}
\def\Ito{It\^o}
\def\bB{\bg{\beta}}
\def\blangle{\bigl\langle}
\def\Blangle{\Bigl\langle}
\def\brangle{\bigr\rangle}
\def\Brangle{\Bigr\rangle}

\def\hatl{\widehat{\lambda}}
\def\part{\partial}
\def\ubar{\underbar}
\def\ul{\underline}
\def\ol{\overline}
\def\vp{\varpi}
\def\nn{\nonumber}
\def\be{\begin{equation}}
\def\ee{\end{equation}}
\def\bea{\begin{eqnarray}}
\def\eea{\end{eqnarray}}
\def\bg{\boldsymbol}
\def\bull{$\bullet~$}

\newcommand{\Slash}[1]{{\ooalign{\hfil/\hfil\crcr$#1$}}}
\vspace{3mm}
\begin{abstract}
In this article, we consider the problem of equilibrium price formation
in an incomplete securities market consisting of one major financial firm and
a large number of minor firms.
They carry out continuous trading via the securities exchange to 
minimize their cost while facing idiosyncratic and common noises as well as stochastic order flows from 
their individual clients.  The equilibrium price process that balances  demand and supply of the securities,
including the functional form of the price impact for the major firm,  is derived endogenously
both in the market of finite population size and in the corresponding mean field limit.
\end{abstract}


\section{Introduction}
In the traditional setups for financial derivatives and portfolio theories,
a security price process is  given exogenously as a part of the model inputs.
On the other hand, in the field of financial economics, 
the problem of equilibrium price formation has been one of the central issues,
which seeks an appropriate price process
that balances demand and supply of securities among a large number of 
agents endogenously based on their  preferences and rational actions.
The intrinsic difficulty for the latter comes from the strategic interactions among the agents.

The progress in the mean field game (MFG) theory in the last decade has opened a new promising 
approach to study the long-standing problem of multi-agent games. 
Since the publication of seminal works by Lasry \& Lions~\cite{Lions-1, Lions-2, Lions-3}
and  Huang, Malhame \& Caines~\cite{Caines-Huang, Caines-Huang-1, Caines-Huang-2, Caines-Huang-3},
which characterizes the Nash equilibrium by a coupled system of Hamilton-Jacobi-Bellman (HJB)
and Kolmogorov equations, 
the mean field game has been one of the central themes among many researchers.

Carmona \& Delarue~\cite{Carmona-Delarue-MFG, Carmona-Delarue-MFTC}
developed a probabilistic approach to the mean field games and 
mean-field type control problems based on a forward-backward 
stochastic differential equation (FBSDE) of McKean-Vlasov type.
Lacker~\cite{Lacker-1, Lacker-2} initiated the weak formulation of the mean field games
by applying the relaxed-control technique. An extension to the so-called 
extended mean field games was recently made by Djete~\cite{Djete1, Djete2}.
The mean field games in the presence of common noise were developed by Carmona et.al.~\cite{Carmona-Delarue-Lacker}
in the framework of weak solutions. 
Lauri\`ere \& Tangpi~\cite{Lauriere} generalized the concept of propagation of chaos
for forward and backward weakly interacting particles.
Since the mean field game theory can decompose a complicated Nash system arising from stochastic differential games
into a separate optimization and an additional fixed point problem, 
it has found vast applications involving many homogeneous agents
competing through symmetric interactions. 
For interested readers, there exist excellent monographs 
such as \cite{Bensoussan-mono, Gomes-eco,  Gomes-reg, Kolokoltsov-mono}
for analytic approach and \cite{Carmona-Delarue-1, Carmona-Delarue-2} for probabilistic approach.
See also the lecture notes by Cardaliaguet \cite{Cardaliaguet-note}.

Since the original MFG setting assumes the homogeneous agents,
one natural extension is to allow multiple types of populations,
where the cost functions
as well as the coefficient functions of the state dynamics can be different 
population by population. See, for example, \cite{Achdou-segregation,Bensoussan-mfg-paper, Cirant, Feleqi, Yu-X}
for analytic approach and \cite{Fujii-mfg} for probabilistic approach.
Another important direction of research is to allow the existence of a major agent 
whose importance does not diminish even in the large population limit of the minor agents.
Huang~\cite{Huang} introduced linear-quadratic mean field games with a major agent, which was 
extended by Nourian \& Caines~\cite{Nourian-Caines} to a general nonlinear dynamical system.
Bensoussan et.al.~\cite{Bensoussan-Major} and Carmona \& Zhu~\cite{Carmona-Zhu}
further developed the framework to allow the major agent to directly influence the law of the minor agents.
The former considered the Stackelberg equilibrium and the latter dealt with the Nash equilibrium.
See also \cite{Firoozi-2} for recent generalization in the linear-quadratic system, 
and \cite{Cardaliaguet, Lasry-Lions} which studies the master equation for the mean field games with a major agent.

These developments of the MFG theory have been successfully applied to various problems regarding
in particular, the energy and financial markets which naturally involve a large number of  agents
with similar preferences. A popular phenomenological approach used to fit to the concept of Nash equilibrium
is to assume that the relevant asset price is decomposed 
into two parts; one is a so-called fundamental price, which is exogenously given and
assumed to be independent of the agents' actions,  and the other part representing
the market friction which is often assumed to be proportional to the average trading speed among the agents.
One can find in \cite{Matoussi, Jaimungal,  Djehiche-E, Feron,  Fu-Horst-1,  Gueant-Oil, Lehalle}
interesting applications  to, optimal trading,  liquidation, energy production, optimal use of smart grids, etc.
In particular, we refer to Fu \& Horst~\cite{Fu-Horst-2},  Evangelista \& Thamsten~\cite{Evangelista} and
F\'eron et.al.\cite{Feron-2} who studied the optimal liquidation and trading problems 
in the mean-field games with a major player.

As for the problem of equilibrium price formation, which requires the prices to balance 
demand and supply of the corresponding assets,  application of the MFG theory has been surprisingly rare.
The first contribution in this direction was made by Gomes \& Saude~\cite{Gomes-Saude}
who modeled the electricity price process using the analytic approach.
Recently, Shrivats et.al.\cite{Firoozi} and Fujii \& Takahashi~\cite{Fujii-Takahashi} 
independently proposed a probabilistic model for equilibrium price formation.
In the former work, the authors studied the solar renewable energy certificate (SREC) market and 
derived the equilibrium SREC price using McKean-Vlasov FBSDEs.
As in \cite{Gomes-Saude}, they assumed that each agent is subject to an independent noise
and applied the fixed-point technique developed by \cite{Carmona-Delarue-MFG} to obtain a deterministic process for the 
equilibrium price. In the latter, we studied the price process of general financial assets using a stylized model of 
the securities exchange. In contrast to \cite{Gomes-Saude, Firoozi}, we  included a common noise 
which affects all the agents. Since the existence of the common noise makes it impossible to 
use the fixed-point technique, we resorted to the continuation method developed by 
Yong~\cite{Yong-C} and Peng \& Wu~\cite{Peng-Wu} 
to solve the conditional McKean-Vlasov FBSDEs directly under the appropriate monotone conditions.
In the accompanying work \cite{Fujii-mfg-convergence}, we proved the strong convergence of 
the finite agent equilibrium to the corresponding mean field limit given in \cite{Fujii-Takahashi}.
Note that, if we only want a short-term solution, the monotone conditions are unnecessary.
In economic terms, they prevent the price bubbles/crashes from happening so that the price process 
is well-posed for an arbitrary interval. Roughly speaking, they require the demand of the 
securities decreases when their prices rise.

In the current paper, we further developed the model studied in the two preceding works \cite{Fujii-Takahashi, Fujii-mfg-convergence}
by including a major agent.
As long as we know, this is the first attempt to solve the problem of equilibrium price formation 
with a major agent under the market-clearing condition.
For a given order flow from the major agent, a properly functioning market
is expected to produce an equilibrium price process as in \cite{Fujii-Takahashi, Fujii-mfg-convergence}.
Since the equilibrium price process of the securities  becomes dependent on the  trading strategy of the major agent,
her optimization problem ends up in minimizing the cost with her own feedback effects into account,
which is given by a large system of controlled-FBSDEs in 
the case of the finite population market, and by controlled-FBSDEs of conditional McKean-Vlasov type in 
the limit of large population size.
In order to guarantee the optimality, we prove the new verification theorem for the controlled-FBSDEs of conditional 
McKean-Vlasov type. Although we are forced to assume a linear-quadratic setup (with stochastic coefficients)
for the minor agents in order to make the verification theorem hold,
we keep a general non-linear cost function for the major agent.
The resultant system of fully-coupled FBSDEs is solved once again by the continuation method.
The equilibrium price process that balances  demand and supply of the securities,
including the functional form of the price impact for the major agent,  is derived endogenously
both in the market of finite population size and in the corresponding mean field limit.
Lastly, we show the strong convergence of the finite agent equilibrium to the corresponding 
mean-field limit. Note that, it is quite rare that one can prove the strong convergence of $N$-agent equilibrium
to the corresponding mean-field limit outside the explicitly solvable linear-quadratic (LQ) settings.
To the best of the authors' knowledge, this is the first example of this kind in the presence of 
a major player in the non-LQ setups.  As an important byproduct, we obtain
the direct estimate on the difference of the equilibrium price between the two markets.

The organization of the paper is as follows:
After explaining the notations in Section 2, we solve the equilibrium price formation 
for the finite population market in Section 3.
The corresponding problem in the mean-field limit is solved in Section 4.
In Section 5, we prove the strong convergence of the finite population equilibrium to the
corresponding mean field limit and give the stability result for the market-clearing price between the two cases.
Section~6 gives a brief discussion on the special case in which the securities have a specified date of maturity  
with exogenously determined payoffs, as in the case for Futures, Bonds and other financial derivatives. 
A general verification theorem for the optimization problem with respect to the controlled-FBSDEs
is provided in Appendix.

\section{Notations}
\label{sec-notation}
We use the same notation adopted in the work~\cite{Fujii-mfg-convergence}.
We introduce (N+1) complete probability spaces:
\bea
(\ol{\Omega}^0, \ol{\calf}^0,\ol{\mbb{P}}^0) \quad {\rm{and}} \quad (\ol{\Omega}^i,\ol{\calf}^i,\ol{\mbb{P}}^i)_{i=1}^N~,\nn
\eea
endowed with filtrations $\ol{\mbb{F}}^i:=(\ol{\calf}_t^i)_{t\geq 0}$, $i\in\{0,\cdots, N\}$.
Here, $\ol{\mbb{F}}^0$ is the completion of the 
filtration generated by $d^0$-dimensional Brownian motion $\bg{W}^0$ (hence right-continuous)
and, for each $i\in\{1,\cdots, N\}$,  $\ol{\mbb{F}}^i$ is the complete and right-continuous augmentation of the filtration
generated by $d$-dimensional Brownian motions $\bg{W}^i$
as well as a $\bg{W}^i$-independent $n$-dimensional  square-integrable random variables $(\xi^i)$. 
We also introduce the product probability spaces
\be
\Omega^i=\ol{\Omega}^0\times \ol{\Omega}^i, \quad \calf^i, \quad \mbb{F}^i=(\calf_t^i)_{t\geq 0}, \quad \mbb{P}^i~, i\in\{1,\cdots, N\}\nn
\ee
where $(\calf^i,\mbb{P}^i)$ is the completion of $(\ol{\calf}^0\otimes \ol{\calf}^i, \ol{\mbb{P}}^0\otimes \ol{\mbb{P}}^i)$
and $\mbb{F}^i$ is the complete and right-continuous augmentation of $(\ol{\calf}_t^0\otimes \ol{\calf}_t^i)_{t\geq 0}$.
In the same way, we  define  the complete probability space $(\Omega,\calf,\mbb{P})$ endowed with $\mbb{F}=(\calf_t)_{t\geq 0}$
satisfying the usual conditions as a product of $(\ol{\Omega}^i,\ol{\calf}^i,\ol{\mbb{P}}^i;\ol{\mbb{F}}^i)_{i=0}^N$. 

Throughout the work, the symbol $L$ and $L_0$ denote given positive constants, 
the symbol $C$ a general positive constant which may change line by line. 
For a given constant $T>0$, we use the following notation for frequently encountered spaces:\\
\bull $\cals^n_+$ denotes the space of $n\times n$ strictly positive definite matrices. \\
\bull $\cals^n$ denotes the space of $n\times n$ positive semidefinite matrices. \\
\bull $\mbb{L}^2(\calg; \mbb{R}^d)$ denotes the set of $\mbb{R}^d$-valued $\calg$-measurable 
square integrable random variables.\\
\bull $\mbb{S}^2(\mbb{G};\mbb{R}^d)$ is the set of $\mbb{R}^d$-valued $\mbb{G}$-adapted continuous processes $\bg{X}$ satisfying
\bea
||X||_{\mbb{S}^2}:=\mbb{E}\bigl[\sup_{t\in[0,T]}|X_t|^2\bigr]^\frac{1}{2}<\infty~. \nn
\eea
\bull $\mbb{H}^2(\mbb{G};\mbb{R}^d)$ is the set of $\mbb{R}^d$-valued $\mbb{G}$-progressively measurable processes $\bg{Z}$ satisfying
\bea
||Z||_{\mbb{H}^2}:=\mbb{E}\Bigl[\Bigl(\int_0^T |Z_t|^2 dt\Bigr)\Bigr]^\frac{1}{2}<\infty~. \nn
\eea
\bull $\call(X)$ denotes the law of a random variable $X$.\\
$\bullet~\calp(\mbb{R}^d)$ is the set of probability measures on $(\mbb{R}^d,\calb(\mbb{R}^d))$. \\
$\bullet~\calp_p(\mbb{R}^d)$ with $p\geq 1$ is the subset of $\calp(\mbb{R}^d)$ with finite $p$-th moment; i.e.,
the set of $\mu\in \calp(\mbb{R}^d)$ satisfying
\bea
M_p(\mu):=\Bigl(\int_{\mbb{R}^d}|x|^p \mu(dx)\Bigr)^\frac{1}{p}<\infty~.\nn
\eea
We always assign $\calp_p(\mbb{R}^d)$ with $(p\geq 1)$ the $p$-Wasserstein distance $W_p$,
which makes  $\calp_p(\mbb{R}^d)$ a complete separable metric space.
It is defined by, for any $\mu, \nu\in \calp_p(\mbb{R}^d)$, 
\bea
W_p(\mu,\nu):={\inf}_{\pi\in\Pi_p(\mu,\nu)}\Bigl[\Bigl(\int_{\mbb{R}^d\times \mbb{R}^d} |x-y|^p \pi(dx,dy)\Bigr)^\frac{1}{p}\Bigr]
\label{def-W}
\eea
where $\Pi_p(\mu,\nu)$ denotes the set of probability measures in $\calp_p (\mbb{R}^d\times \mbb{R}^d)$
with marginals $\mu$ and $\nu$. For more details,  see Chapter 5 in \cite{Carmona-Delarue-1}. \\
\bull For any $N$ variables $(x^i)_{i=1}^N$, we write its empirical mean as
\be
\mg{m}((x)):=\mg{m}((x^i)_{i=1}^N):=\frac{1}{N}\sum_{i=1}^N x^i.\nn
\ee
We frequently omit the arguments such as $(\mbb{G},\mbb{R}^d)$ in the above definitions when there is no confusion 
from the context. 

\section{Equilibrium of finite population size}
\label{sec-finite-agent}
\subsection{Problem description}
In the preceding works, we have been interested in the equilibrium price formation in a financial market
among a large number of security firms. 
Every firm (agent) is supposed to have many individual clients
who cannot directly access to the exchange. Therefore,  
every agent supposed to face the stochastic order flows from his individual clients in addition to the 
idiosyncratic as well as common market shocks.  Under such an environment,
they carry out optimal trading  via the common  exchange 
to minimize their cost functions. Importantly, since there exist very large number of agents,
every agent considers that his market share is negligibly small and hence
that there is no direct market impact from his trading. In other words, they 
behave as {\it price takers}. The problem of equilibrium price formation is to search an appropriate price process of securities
which equalize the demand and supply based on the agents' cost functions
and the state dynamics.  In the presence of common shocks, the 
price process inevitably becomes stochastic.
Such a problem has been investigated in our two preceding papers~\cite{Fujii-Takahashi, Fujii-mfg-convergence},
where the former treats the mean-field limit and the latter proves
the strong convergence to the mean-field limit from the corresponding equilibrium of finite population.

The new twist in the current paper is the presence of one major agent, a huge financial firm, 
who knows that her trading volume has a significant market share.
For a given order flow from the major agent, a properly functioning market
is expected to produce an equilibrium price process so that it matches the net demand 
and supply among all the agents.
Through this function of the market, the equilibrium price process of the securities 
becomes dependent on the trading strategy of the major agent.
Therefore, 
her optimization problem  
ends up in minimizing the cost with her own feedback effects into account.
We then finally obtain the market equilibrium price process  by solving the major agent's optimal strategy.
In the following, we first solve this problem in the market with finite population size.
The minor agents are allowed to be heterogeneous so that the coefficients functions for their state processes
as well as the cost functions can be different from each other.  The large population limit of minor agents will be studied in 
later sections. \\ 

Let us now describe the setup more concretely. 
There are $N$ minor agents indexed by $i=1,\cdots, N$. The major agent is always labeled by the index $0$.
The number of securities traded in the market is assumed to be $n\in \mbb{N}$.
Each minor agent $i\in\{1,\cdots, N\}$ tries to solve the cost minimization problem
among the admissible strategies  $\mbb{A}^i:=\mbb{H}^2(\mbb{F};\mbb{R}^n)$
\be
\inf_{\bg{\alpha}^i \in \mbb{A}^i} J^i(\bg{\alpha}^i)
\label{problem-minor}
\ee
with some functions $f_i$ and $g_i$, which denotes the running as well as terminal costs, respectively:
\bea
J^i(\bg{\alpha}^i):=\mbb{E}\Bigl[\int_0^T f_i(t,X_t^i, \alpha_t^i, \vp_t, \Lambda_t, c_t^0, c_t^i)dt+g_i(X_T^i, \vp_T, c_T^0, c_T^i)\Bigr]~. \nn
\eea
The dynamic  constraint, which is the time evolution of the securities' position size of the $i$th agent, is given by
\bea
dX_t^i=\bigl(\alpha_t^i+l_i(t,c_t^0,c_t^i)\bigr)dt+\sigma_i^0(t,c_t^0,c_t^i)dW_t^0+\sigma_i (t,c_t^0, c_t^i)dW_t^i, \quad t\in[0,T] \nn
\eea 
with $X_0^i=\xi^i$.
Here, $\xi^i\in \mbb{L}^2(\ol{\calf}_0^i; \mbb{R}^n)$ denotes
the size of the initial position, which is assumed to have the common law for every $1\leq i\leq N$.
$(\vp_t)_{t\in[0,T]} \in \mbb{H}^2(\mbb{F};\mbb{R}^n)$ denotes the market price process of the $n$ securities.
In the end, we want to determine $(\vp_t)_{t\in[0,T]}$ endogenously so that it equalizes the 
amount of demand and supply. 
$(c_t^0)_{t\geq 0}\in \mbb{H}^2(\ol{\mbb{F}}^0;\mbb{R}^n)$ with $c_T^0\in \mbb{L}^2(\ol{\calf}_T^0;\mbb{R}^n)$
denotes the coupon payments from the securities or the market news affecting  all the agents, 
while $(c_t^i)_{t\geq 0}\in \mbb{H}^2(\ol{\mbb{F}}^i;\mbb{R}^n)$ with $c_T^i\in \mbb{L}^2(\ol{\calf}_T^i;\mbb{R}^n)$ 
denotes some idiosyncratic shocks affecting only the $i$th agent.
Moreover,  $(c_t^i)_{t\geq 0}$ are also assumed to have the common law for all $1\leq i\leq N$.
$(\Lambda_t)_{t\in[0,T]}$ is an $\ol{\mbb{F}}^0$-adapted process related to the trading fee to be paid to the exchange.
The terms involving $(l_i, \sigma_i^0, \sigma_i)$ denote the order flow to the $i$th agent from his individual clients
through the over-the-counter (OTC) market. Each minor agent controls $(\alpha_t^i)_{t\in[0,T]}$,
which is an $\mbb{R}^n$-valued process denoting the trading speed of the $n$ securities via the exchange.
More precisely,  $(\alpha^i_t)^k dt$, $1\leq k\leq n$, denotes the number of shares of the $k$th security
bought (or sold if negative) within the time interval $[t,t+dt]$ by the $i$th agent.
Note that, in addition to the random initial states $(\xi^i)_{i=1}^N$, we have $d_0$-dimensional common noise $W^0$
and $N$ $d$-dimensional idiosyncratic noises $(W^i)_{i=1}^N$. Since we impose no restriction 
on the size among $(n, d_0, d, N)$, we have an incomplete securities market in general.
For more information, see \cite[Section 3]{Fujii-Takahashi},
which explains the financial interpretation of each term in details.

When the number of agents $N$ is sufficiently large, it is natural to 
assume that each minor agent consider himself as a {\it price taker}.
Throughout the paper, we assume that this is the case.
This means that each minor agent  tries to solve the optimization problem by treating $(\vp_t)_{t\geq 0}$ as an exogenous 
process.    Suppose that the trading strategy of the major agent is given by $(\beta_t)_{t \in[0,T]}$,  
which denotes her trading speed.
For  given order flow  $(\beta_t)_{t\in[0,T]}$,
the financial market is expected to produce an equilibrium price process $(\vp_t)_{t\in[0,T]}$ 
which equalizes the demand and supply among all the agents.
Our first goal is to find such a price process $(\vp_t)_{t\in[0,T]}$ which achieves
\be
\sum_{i=1}^N  \wh{\alpha}_t^i+\beta_t=0
\ee
$dt\otimes d\mbb{P}$-a.e.,  where $\bigl((\wh{\alpha}_t^i)_{t\in[0,T]}\bigr)_{i=1}^N$
are the optimal trading strategies of the minor agents solving $(\ref{problem-minor})$ 
based on this price process $(\vp_t)_{t\in[0,T]}$.

We shall show that the resultant equilibrium price process becomes dependent on $(\beta_t)_{t\in[0,T]}$
i.e., we have $\bigl(\vp_t(\bg{\beta})\bigr)_{t\in[0,T]}$.
Note that the minor agents do not {\it directly} care about $(\beta_t)_{t\in[0,T]}$.
They  are just destined to face,  as price takers,  the exogenous market price process, which  
happens to depend on the major's strategy when it clears the market.
The problem of the major agent is now to solve
\be
\inf_{\bg{\beta}\in \mbb{A}^0}J^0(\bg{\beta}) 
\label{problem-major}
\ee
with the cost functional depending on $f_0^{(N)}$ and $g_0^{(N)}$:
\bea
J^0(\bg{\beta}):=\mbb{E}\Bigl[\int_0^T f_0^{(N)}(t,X_t^0,\beta_t, \vp_t(\bg{\beta}), \Lambda_t^0, c_t^0)dt+
g_0^{(N)}(X_T^0, c_T^0)\Bigr], \nn
\eea
with her own feedback effects taken into account. $(\Lambda_t^0)_{t\in[0,T]}$
is an $\ol{\mbb{F}}^0$-adapted process related to the trading fee to be paid to the exchange.
The state dynamics of the major agent describing her position size is assumed to follow
\bea
dX_t^0=\bigl(\beta_t+l_0^{(N)}(t,c_t^0)\bigr)dt+\sigma_0^{(N)}(t,c_t^0)dW_t^0, \quad t\in[0,T]
\eea
with some initial condition $X_0^0\in \mbb{R}^n$.
The superscript $(N)$ of the coefficient functions
is added to indicate that there are $(N)$ minor agents. It becomes useful when we take 
the large-$N$ limit in later sections.
We assume that the space of admissible strategies for the major agent is  given by
$\mbb{A}^0:=\mbb{H}^2(\mbb{F};\mbb{R}^n)\cap\{\beta_T=0\}$, where the constraint $\beta_T=0$ is 
added in order to forbid the last-time price manipulation. 

In our framework, the price process including the feedback effects from the major's action
is determined endogenously. This is a clear contrast to the existing literature
dealing with the optimal execution strategy,
where the form of the price impact as well as the fundamental price process are exogenously given.
With appropriate modifications of the cost functions 
and their interpretations,  the current setup may be useful also for economic analysis, for example,
 the market  involving one major producer and a large number of small consumers.

Before going to the details, let us comment on the information structure for the agents. 
\begin{remark}
\label{remark-information-structure}
If possible, we naturally want to restrict the space of admissible strategies 
for each minor agent to $\mbb{A}^i=\mbb{H}^2(\mbb{F}^i; \mbb{R}^n),~1\leq i\leq N$
and that for the major agent to $\mbb{A}^0=\mbb{H}^2(\ol{\mbb{F}}^0; \mbb{R}^n)\cap \{\beta_T=0\}$.
In other words, we want to realize a market in which each agent only cares about the common market shocks adapted to $\ol{\mbb{F}}^0$
and his/her own idiosyncratic shocks adapted to $\ol{\mbb{F}}^i$. This would be a much plausible model
for the real financial market than our setup given above. Unfortunately,  this looks impossible in the market consisting of finite number of agents 
since, in general,  the market-clearing price  does not solely adapted to $\ol{\mbb{F}}^0$ but is dependent on the idiosyncratic shocks, too.

As already observed in \cite{Fujii-Takahashi, Fujii-mfg-convergence}, 
we shall see that this ideal situation is actually realized in the large population limit.
There, we can restrict the admissible strategy of the $i$th minor agent to $\mbb{A}^i=\mbb{H}^2(\mbb{F}^i;\mbb{R}^n)$,
and that of the major agent to $\mbb{A}^0=\mbb{H}^2(\ol{\mbb{F}}^0;\mbb{R}^n)\cap\{\beta_T=0\}$.
In fact, we can find $(\vp_t)_{t\in[0,T]}$ is an $\ol{\mbb{F}}^0$-adapted process, i.e. the market-clearing price  
is dependent only on the common market shocks. By the convergence analysis, we shall see
that this is approximately true when the population size is large enough.
\end{remark}

\subsection{Solving the problem for the minor agents}
\label{sec-minor-problem}
Let us solve the problem for each minor agent with  given order flow $(\beta_t)_{t\in[0,T]}\in \mbb{A}^0$
of the major agent. 
This is done in a completely parallel manner with our previous work~\cite{Fujii-mfg-convergence}.
We first specify the details of the functions introduced in the last section.
For each $1\leq i\leq N$, we consider the following measurable functions:
\bea
&&(l_i, \sigma_i^0, \sigma_i) :[0,T]\times \mbb{R}^n\times \mbb{R}^n \ni (t,c^0, c^i)\nn \\
&&\qquad\qquad  \mapsto (l_i(t,c^0,c^i), \sigma_i^0(t,c^0,c^i), \sigma_i(t,c^0,c^i)) \in (\mbb{R}^n, \mbb{R}^{n\times d_0},
\mbb{R}^{n\times d}), \nn \\
&&\ol{f}_i:[0,T]\times (\mbb{R}^n)^4\ni (t,x,\vp, c^0,c^i)\mapsto \ol{f}_i(t,x,\vp, c^0,c^i)\in \mbb{R}, \nn   \\
&&\ol{g}_i:(\mbb{R}^n)^3\ni (x,c^0,c^i)\mapsto \ol{g}_i(x,c^0,c^i)\in \mbb{R}, \nn
\eea
as well as $f_i:[0,T]\times (\mbb{R}^n)^3\times \cals^n_+ \times (\mbb{R}^n)^2\rightarrow \mbb{R}$
 and $g_i:(\mbb{R}^n)^4\rightarrow \mbb{R}$ defined by
\bea
&&f_i(t,x,\alpha,\vp, \Lambda, c^0,c^i):=\langle \vp, \alpha\rangle+\frac{1}{2}\langle \alpha, \Lambda \alpha\rangle+\ol{f}_i(t,x,\vp,c^0,c^i), \nn \\
&&g_i(x,\vp,c^0,c^i):=-\del \langle \vp, x\rangle+\ol{g}_i(x,c^0,c^i). \nn
\eea

Let us explain the economic meaning of the cost functions.
By buying (or selling if negative) with speed $\alpha_t$, each agent pays (or receives if negative)
$\langle \alpha_t, \vp_t\rangle dt$ amount of cash in the time interval $[t,t+dt]$.
In addition to this direct cost,  we suppose that each agent has to pay the service fees 
to the securities exchange $\frac{1}{2}\langle \alpha_t, \Lambda \alpha_t\rangle dt$ where
$\Lambda$ is an $n\times n$ positive definite matrix. 
These costs are represented by the first two terms of the function $f_i$.
The first term of $g_i$ denotes
the mark-to-market value at the closing time with some discount factor $\del\in [0,1)$.\footnote{We shall see that the condition $\del <1$
is necessary to obtain well-defined terminal condition for the equilibrium.}
The above three terms are assumed to be common across the agents 
since there is no strong motivation to suppose otherwise.
The remaining terms represented by functions $\ol{f}_i$ and $\ol{g}_i$
can be used to distinguish various characters among the agents.
The function $\ol{f}_i$ is supposed to represent the  running costs
which can be  dependent on the position size, cash flows, prices of the securities
as well as any relevant news available to each agent. 
The function $\ol{g}_i$ puts some penalty on the position size at the terminal time $T$.
In particular,  we can make the $i$th agent more risk averse
by assigning stronger convexity on $x$ for $\ol{f}_i$ and/or $\ol{g}_i$.

\begin{example}
Suppose that the $n$ securities have continuous dividend payments $(c_t^0)_{t\in[0,T)}$ as well as
the rump-sum payment $c_T^0$ at time $T$. In this case,  it may be natural to consider
\be
\begin{split}
\ol{f}_i(t,x,\vp,c^0,c^i)&=-\langle c^0, x\rangle+\ol{f}_i^\prime (t,x,\vp,c), \\
\ol{g}_i(x,\vp,c^0,c^i)&=-\langle c^0, x\rangle+\ol{g}_i^\prime(x,\vp,c), 
\end{split} 
\nn
\ee
with some appropriate measurable functions $\ol{f}_i^\prime$ and $\ol{g}_i^\prime$. 
Here, the first term $\langle c^0,x\rangle$ denotes the benefit from the receipt of the cash flow.
The idiosyncratic shock $c^i$ can be used in various ways.
For example, we may used it to change the risk-averseness with respect to the position size $x$ of each agent according to the 
arrival of the idiosyncratic information. 
In fact, this is represented by the functions $c^f_i(\cdot, c^i)$ and $c_i^g(\cdot, c^i)$ explained in {\rm (iv)}
of Assumption~\ref{assumption-Minor-A} given below.
\end{example}

Let us also introduce the following measurable functions $(c_i^f, c_i^g, h_i^f, h_i^g)$ for each $1\leq i\leq N$:
\bea
&&c^f_i:[0,T]\times (\mbb{R}^n)^2\ni (t,c^0,c^i)\mapsto c^f_i(t,c^0,c^i)\in \cals^n_+, \nn \\
&&c^g_i:(\mbb{R}^n)^2\ni(c^0,c^i)\mapsto c^g_i(c^0,c^i)\in \cals^n_+, \nn \\
&&h_i^f:[0,T]\times (\mbb{R}^n)^2\mapsto h_i^f(t,c^0,c^i)\in \mbb{R}^n, \nn \\
&&h_i^g: (\mbb{R}^n)^2\mapsto h_i^g(c^0,c^i)\in \mbb{R}^n.\nn
\eea

We assume the following conditions:
\begin{assumption}{\rm (Minor-A)}
\label{assumption-Minor-A}
 Uniformly in $1\leq i\leq N$, the functions satisfy the followings:\\
{\rm (i)} $(\Lambda_t)_{t\in[0,T]}$ is an $\ol{\mbb{F}}^0$-progressively measurable 
$\cals^n_+$-valued process such that there exist some positive constants $0<\ul{\lambda}\leq \ol{\lambda}<\infty$
satisfying $\ul{\lambda}|\theta|^2\leq \langle \theta, \Lambda_t \theta\rangle\leq \ol{\lambda}|\theta|^2$
for every $(\omega, t, \theta)\in \Omega\times [0,T]\times \mbb{R}^n$. \\
{\rm (ii)} For any $(t,c^0,c^i)\in [0,T]\times (\mbb{R}^n)^2$,  
\be
|l_i(t,c^0,c^i)|+|\sigma_i^0(t,c^0,c^i)|+|\sigma_i(t,c^0,c^i)|\leq L(1+|c^0|+|c^i|). \nn
\ee
{\rm (iii)} For any $(t,x, \vp,c^0,c^i)\in [0,T]\times (\mbb{R}^n)^4$, 
\bea
|\ol{f}_i(t,x,\vp,c^0,c^i)|+|\ol{g}_i(x,c^0,c^i)|\leq L(1+|x|^2+|\vp|^2+|c^0|^2+|c^i|^2). \nn
\eea
{\rm (iv)} For any $(t,x,\vp, c^0,c^i)\in[0,T]\times (\mbb{R}^n)^4$, $\ol{f}_i$ and $\ol{g}_i$ are once continuously differentiable in $x$
with $\vp$-independent derivatives, and the functions $\part_x \ol{f}_i$ and $\part_x \ol{g}_i$
have the following affine-form in $x$:
\bea
&&\part_x\ol{f}_i(t,x,\vp, c^0,c^i)~\bigl(=:\part_x\ol{f}_i(t,x,c^0,c^i)\bigr)=c_i^f(t,c^0,c^i)x+h^f_i(t,c^0,c^i), \nn \\
&&\part_x \ol{g}_i(x,c^0,c^i)=c_i^g(c^0,c^i)x+h_i^g(c^0,c^i). \nn
\eea
Moreover, the functions $(c_i^f, c_i^g, h_i^f, h_i^g)$ satisfy
\bea
&&|h_i^f(t,c^0,c^i)|+|h_i^g(c^0,c^i)|\leq L(1+|c^0|+|c^i|), \nn \\
&&|c_i^f(t,c^0,c^i)|+|c_i^g(c^0,c^i)|\leq L, \nn \\
&&\blangle \theta, c_i^f(t,c^0,c^i)\theta\brangle \geq \gamma^f |\theta|^2, \quad \blangle \theta, c_i^g(c^0,c^i)\theta\brangle
\geq \gamma^g |\theta|^2, \quad \forall \theta\in\mbb{R}^n, \nn
\eea
with some positive constants $\gamma^f, \gamma^g>0$.
\end{assumption}
This is a special situation studied in Section 3.1 of \cite{Fujii-mfg-convergence}. 
In fact, the conditions in Assumption (Minor-A) are significantly more stringent than those used in \cite{Fujii-mfg-convergence}.
We do this in order to avoid introducing many sets of assumptions incrementally
in later sections. In particular, the affine-form condition in {\rm (iv)}
is to be used when we verify the optimality condition for the major agent based on  Theorem~\ref{th-A-verification}.
The associated (reduced) Hamiltonian for the $i$th agent 
$H_i:[0,T]\times (\mbb{R}^n)^4\times \cals_+^n \times (\mbb{R}^n)^2\rightarrow \mbb{R}$
is given by
\bea
H_i(t,x,y,\alpha, \vp, \Lambda, c^0,c^i):=\blangle y, \alpha+l_i(t,c^0,c^i)\brangle+f_i(t,x,\alpha, \vp, \Lambda, c^0,c^i), \nn
\eea
which is jointly convex in $(x,y,\alpha)$ and strictly so in $(x,\alpha)$.
The unique minimizer $\alpha$ of $H_i$ is  given by
\be
\wh{\alpha}(y,\vp):=-\ol{\Lambda}(y+\vp), \nn
\ee
with $\ol{\Lambda}:=\Lambda^{-1}$.
Therefore, the adjoint equation associated with the problem $(\ref{problem-minor})$ for the $i$th agent 
arising from the stochastic maximum principle is given by, for $t\in[0,T]$, 
\bea
\begin{cases}
dX_t^i=\bigl(-\ol{\Lambda}_t(Y_t^i+\vp_t)+l_i(t,c_t^0,c_t^i)\bigr)dt+\sigma_i^0(t,c_t^0,c_t^i)dW_t^0+
\sigma_i(t,c_t^0,c_t^i)dW_t^i, \\
dY_t^i=-\part_x \ol{f}_i(t,X_t^i, c_t^0,c_t^i)dt+Z_t^{i,0}dW_t^0+\sum_{j=1}^N Z_t^{i,j}dW_t^j, 
\end{cases}
\label{eq-adjoint-minor}
\eea
with $X_0^i=\xi^i$ and $Y_T^i=-\del \vp_T+\part_x \ol{g}_i(X_T^i,c_T^0,c_T^i)$.

\begin{theorem}
\label{th-optimal-minor}
Let Assumption (Minor-A) be in force. Then, for any $(\vp_t)_{t\in[0,T]}\in \mbb{H}^2(\mbb{F};\mbb{R}^n)$
satisfying $\vp_T\in \mbb{L}^2(\calf_T;\mbb{R}^n)$,  the problem $(\ref{problem-minor})$ for each agent $1\leq i\leq N$ is uniquely 
characterized by the FBSDE $(\ref{eq-adjoint-minor})$ which is strongly solvable 
with a unique solution $(X^i, Y^i, Z^{i,0}, (Z^{i,j})_{j=1}^N)\in \mbb{S}^2(\mbb{F};\mbb{R}^n)\times \mbb{S}^2(\mbb{F};\mbb{R}^n)
\times \mbb{H}^2(\mbb{F};\mbb{R}^{n\times d_0})\times (\mbb{H}^2(\mbb{F};\mbb{R}^{n\times d}))^N$. 
\begin{proof}
This is the direct result of Theorem~3.1 in \cite{Fujii-mfg-convergence}.
One can easily check Assumption~3.1 in \cite{Fujii-mfg-convergence} is satisfied under (Minor-A).
Although $(\Lambda_t)_{t\in[0,T]}$ is now stochastic, it does not introduce any additional difficulty.
The existence of the unique solution to the FBSDE $(\ref{eq-adjoint-minor})$ 
can also be proved by the direct application of Theorem~2.6 in \cite{Peng-Wu} (with $\beta_1, \mu_1>0$),
which is repeatedly used in the following sections.
\end{proof}
\end{theorem}

\subsection{Deriving the equilibrium price process for a given $(\beta_t)_{t\in[0,T]}$}
From Theorem~\ref{th-optimal-minor}, 
we find that the optimal trading speed of each minor agent $1\leq i\leq N$ is given by
\bea
\wh{\alpha}_t^i=-\ol{\Lambda}_t (Y_t^i+\vp_t), \qquad t\in[0,T],\nn
\eea
for any exogenous input $(\vp_t)_{t\in[0,T]}$. Since the market-clearing condition requires
$\sum_{i=1}^N \wh{\alpha}_t^i+\beta_t=0$, $dt\otimes d\mbb{P}$-a.e. 
the market price process needs to satisfy
\bea
\vp_t=-\mg{m}\bigl((Y_t)\bigr)+\Lambda_t \frac{\beta_t}{N}, \quad t\in[0,T]. 
\label{eq-price-beta}
\eea
This relation suggests a large system of fully-coupled FBSDEs given below:
for $1\leq i\leq N$, 
\bea
\begin{cases}
\display dX_t^i=\Bigl\{-\ol{\Lambda}_t\bigl(Y_t^i-\mg{m}((Y_t))\bigr)-\frac{\beta_t}{N}+l_i(t,c_t^0,c_t^i)\Bigr\}dt+
\sigma_i^0(t,c_t^0,c_t^i)dW_t^0+\sigma_i(t,c_t^0,c_t^i)dW_t^i, \\
dY_t^i=-\part_x \ol{f}_i(t,X_t^i, c_t^0,c_t^i)dt+Z_t^{i,0}dW_t^0+\sum_{j=1}^N Z_t^{i,j}dW_t^j, 
\end{cases}
\label{eq-minor-clearing}
\eea 
with 
\bea
\label{eq-minor-clearing-terminal}
\begin{cases}
X_0^i=\xi^i,  \\
Y_T^i=\frac{\del}{1-\del}\mg{m}\Bigl(\bigl(c_j^g(c_T^0,c_T^j)X_T^j+h_j^g(c_T^0,c_T^j)\bigr)_{j=1}^N\Bigr)+c_i^g(c_T^0,c_T^i)X_T^i+h_i^g(c_T^0,c_T^i)~.
\end{cases}
\eea
The terminal condition for $Y^i$ is implied from 
\bea
Y_T^i=-\del \vp_T+\part_x\ol{g}_i(X_T^i, c_T^0,c_T^i) \nn
\eea
and the fact that $\vp_T=-\mg{m}((Y_T))$ (note that $\beta_T=0$).
We have the following result.

\begin{theorem}
Let Assumption (Minor-A) be in force. With a given strategy $(\beta_t)_{t\in[0,T]}\in \mbb{A}^0$ of the major agent, 
the market-clearing equilibrium with a square integrable price 
process  $(\vp_t)_{t\in[0,T]}\in \mbb{H}^2(\mbb{F};\mbb{R}^n)$ with $\vp_T\in \mbb{L}^2(\calf_T;\mbb{R}^n)$
exists if and only if 
there exists a solution $(X^i, Y^i, Z^{i,0}, (Z^{i,j})_{j=1}^N)\in \mbb{S}^2(\mbb{F};\mbb{R}^n)\times \mbb{S}^2(\mbb{F};\mbb{R}^n)
\times \mbb{H}^2(\mbb{F};\mbb{R}^{n\times d_0})\times (\mbb{H}^2(\mbb{F};\mbb{R}^{n\times d}))^N$, $1\leq i\leq N$
to the $N$-coupled system of FBSDEs $(\ref{eq-minor-clearing})$ with $(\ref{eq-minor-clearing-terminal})$.
\begin{proof}
This is a simple modification of \cite[Theorem 3.2]{Fujii-mfg-convergence}.
The necessity is obvious from Theorem~\ref{th-optimal-minor} and the above discussion.
On the other hand, suppose that there exists a square integrable solution 
to the $N$-coupled FBSDEs $(\ref{eq-minor-clearing})$ with $(\ref{eq-minor-clearing-terminal})$.
Let us define the price process $\vp$ by $(\ref{eq-price-beta})$ using the solution $(Y^i)_{i=1}^N$.
Then, with this $\vp$ as an input, the solution $(y_t^i)_{t\in[0,T]}$ to $(\ref{eq-adjoint-minor})$,
which corresponds to the problem for the $i$th agent,  actually satisfies $y^i=Y^i$ in $\mbb{S}^2(\mbb{F};\mbb{R}^n)$ 
due to the uniqueness of the solution to $(\ref{eq-adjoint-minor})$. Therefore, the market-clearing condition is satisfied.
\end{proof}
\label{th-minor-equilibrium}
\end{theorem}

\begin{assumption}{\rm (Minor-B)}\\
There exists some $\calf_T$-measurable $\cals^n$-valued random variable $\mg{c}$ such that
\bea
\mg{a}:=\frac{\del}{1-\del}||\mg{c}-c_i^g(c^0_T, c_T^i)||_{\infty}<\gamma^g, \quad 1\leq i\leq N. \nn
\eea
\end{assumption}

\begin{theorem}
\label{th-minor-clearing-existence}
Let Assumptions (Minor-A, B) be in force.
Then, for any given $(\beta_t)_{t\in[0,T]}\in \mbb{A}^0$, 
the $N$-coupled system of FBSDEs $(\ref{eq-minor-clearing})$ with $(\ref{eq-minor-clearing-terminal})$
has a unique strong solution $(X^i, Y^i, Z^{i,0}, (Z^{i,j})_{j=1}^N)\in\mbb{S}^2(\mbb{F};\mbb{R}^n)\times 
\mbb{S}^2(\mbb{F};\mbb{R}^n)\times \mbb{H}^2(\mbb{F};\mbb{R}^{n\times d_0})\times (\mbb{H}^2(\mbb{F};\mbb{R}^{n\times d}))^N$,
$1\leq i\leq N$.
\begin{proof}
Let $x^i, y^i\in \mbb{R}^n$ be arbitrary constants.
For notational simplicity, we write $x=(x^i)_{i=1}^N$ and $y=(y^i)_{i=1}^N$. Put
\bea
&&{\rm drift}[x^i](t, y):=-\ol{\Lambda}_t\bigl(y^i-\mg{m}((y))\bigr)-\frac{\beta_t}{N}+l_i(t,c_t^0,c_t^i), \nn \\
&&{\rm drift}[y^i](t, x):=-\part_x \ol{f}_i(t,x^i,c_t^0,c_t^i), \nn \\
&&{\rm terminal}[y^i](x):=\frac{\del}{1-\del}\mg{m}\bigl((c^g_j(c_T^0,c_T^j)x^j+h_j^g(c_T^0,c_T^j))_{j=1}^N\bigr)+c_i^g(c_T^0,c_T^i)x^i+h_i^g(c_T^0,c_T^i). \nn
\eea
For two inputs $(x,y)$ and $(x^\prime, y\pp)$, with the conventions
$\Del x^i:=x^i-x^{i\prime}$, $\Del y^i:=y^i-y^{i\prime}$,
\bea
&&\Del {\rm drift}[x^i](t):={\rm drift}[x^i](t,y)-{\rm drift}[x^i](t,y\pp), \nn \\
&&\Del {\rm drift}[y^i](t):={\rm drift}[y^i](t,x)-{\rm drift}[y^i](t,x\pp), \nn \\
&&\Del {\rm terminal}[y^i]:={\rm terminal}[y^i](x)-{\rm terminal}[y^i](x\pp),  \nn
\eea
we  have
\bea
&&\sum_{i=1}^N \blangle \Del {\rm drift}[x^i](t), \Del y^i\brangle =-\sum_{i=1}^N \blangle \ol{\Lambda}_t \Del y^i, \Del y^i\brangle
+N\blangle \ol{\Lambda}_t\mg{m}((\Del y)), \mg{m}((\Del y))\brangle \leq 0,\nn \\
&&\sum_{i=1}^N \blangle \Del {\rm drift}[y^i](t), \Del x^i\brangle=-\blangle c_i^f(t,c^0_t,c_t^i)\Del x^i, \Del x^i\brangle
\leq -\gamma^f \sum_{i=1}^N |\Del x^i|^2, \nn
\eea
\bea
&&\sum_{i=1}^N \blangle \Del {\rm terminal}[y^i], \Del x^i\brangle\nn  \\
&&\quad=\frac{\del N}{1-\del}\blangle \mg{m}((c_i^g(c_T^0,c_T^i)\Del x^i)_{i=1}^N), \mg{m}((\Del x))\brangle+
\sum_{i=1}^N \blangle c_i^g(c_T^0,c_T^i)\Del x^i, \Del x^i\brangle \nn \\
&&\quad\geq \frac{\del N}{1-\del}\blangle \mg{c}~\mg{m}((\Del x)), \mg{m}((\Del x))\brangle+(\gamma^g-\mg{a})\sum_{i=1}^N |\Del x^i|^2
 \geq (\gamma^g-\mg{a})\sum_{i=1}^N |\Del x^i|^2. 
\label{eq-terminal-cal}
\eea
Thus we can apply Theorem~2.6 in \cite{Peng-Wu} with $(\beta_1, \mu_1)=(\gamma^f, \gamma^g-\mg{a})$
and $G=I$.  See also the proof for Theorem~3.3 in \cite{Fujii-mfg-convergence},
which can be applied in  essentially the same way for the current problem.
\end{proof}
\end{theorem}

\subsection{Optimization problem for the major agent}
We now investigate the optimization problem for the major agent.
From Theorems~\ref{th-minor-equilibrium} and \ref{th-minor-clearing-existence},
her problem is given by $\inf_{\bg{\beta}\in \mbb{A}^0}J^0(\bg{\beta})$
with
\bea
J^0(\bg{\beta}):=\mbb{E}\Bigl[\int_0^T f_0^{(N)}\Bigl(t,X_t^0,\beta_t, -\mg{m}((Y_t))+\Lambda_t\frac{\beta_t}{N}, \Lambda_t^0, c_t^0\Bigr)dt+
g_0^{(N)}(X_T^0, c_T^0)\Bigr], \nn
\eea
subject to the dynamic constraints with $1\leq i\leq N$:
\bea
\begin{cases}
dX_t^0=\bigl(\beta_t+l_0^{(N)}(t,c_t^0)\bigr)dt+\sigma_0^{(N)}(t,c_t^0)dW_t^0,  \\
\display dX_t^i=\Bigl\{-\ol{\Lambda}_t\bigl(Y_t^i-\mg{m}((Y_t))\bigr)-\frac{\beta_t}{N}+l_i(t,c_t^0,c_t^i)\Bigr\}dt+
\sigma_i^0(t,c_t^0,c_t^i)dW_t^0+\sigma_i(t,c_t^0,c_t^i)dW_t^i, \\
dY_t^i=-\part_x \ol{f}_i(t,X_t^i, c_t^0,c_t^i)dt+Z_t^{i,0}dW_t^0+\sum_{j=1}^N Z_t^{i,j}dW_t^j, \quad t\in[0,T]
\end{cases}
\label{eq-state-major}
\eea
with
\bea
\begin{cases}
X_0^0=N\chi^0, \quad \chi^0 \in \mbb{R}^n,  \\
X_0^i=\xi^i,  \\
Y_T^i=\frac{\del}{1-\del}\mg{m}\Bigl(\bigl(c_j^g(c_T^0,c_T^j)X_T^j+h_j^g(c_T^0,c_T^j)\bigr)_{j=1}^N\Bigr)+c_i^g(c_T^0,c_T^i)X_T^i+h_i^g(c_T^0,c_T^i).
\end{cases}
\label{eq-state-terminal-major}
\eea
As we can see, the problem for the major agent turns out to be an optimization with respect to the system of controlled-FBSDEs
instead of controlled-SDEs. See, for relevant information, Appendix~\ref{sec-controlled-fbsde} and the references therein.

\begin{remark}
At first glance,  it may seem to be a linear price impact model popular in the literature dealing with the optimal execution problem.
However,  notice that the term $-\mg{m}((Y_t^i))$ is also dependent on the major agent's strategy in a complicated fashion.
\end{remark}

Since we want to study the large population limit $N\rightarrow \infty$ in later sections,
it is convenient to define the normalized measurable functions:
\bea
&&(\mg{l}_0, \mg{s}_0):[0,T]\times \mbb{R}^n\ni (t,c^0)\mapsto (\mg{l}_0(t,c^0), \mg{s}_0(t,c^0))\in (\mbb{R}^n,\mbb{R}^{n\times d_0}), \nn \\
&&\ol{\mg{f}}_0:[0,T]\times (\mbb{R}^n)^2\ni (t,x,c^0)\mapsto \ol{\mg{f}}_0(t,x,c^0)\in \mbb{R}, \nn \\
&&\mg{g}_0:(\mbb{R}^n)^2\ni (x,c^0)\mapsto \mg{g}_0(x,c^0)\in \mbb{R}, \nn
\eea
and $\mg{f}_0:[0,T]\times (\mbb{R}^n)^3\times \cals^n \times \mbb{R}^n\rightarrow \mbb{R}$ by
\be
\mg{f}_0(t,x,\beta,\vp, \Lambda^0, c^0):=\langle\beta,\vp\rangle+\frac{1}{2}\blangle \beta, \Lambda^0\beta\brangle+
\ol{\mg{f}}_0(t,x,c^0).\nn
\ee
We then define the unnormalized functions by
\be
\begin{split}
&l_0^{(N)}(t,c^0):=N\mg{l}_0(t,c^0), \\
&\sigma_0^{(N)}(t,c^0):=N\mg{s}_0(t,c^0),  \\
&f_0^{(N)}(t,x,\beta,\vp,\Lambda^0,c^0):=N\mg{f}_0\bigl(t,x/N, \beta/N, \vp, \Lambda^0, c^0\bigr), \\
&\ol{f}_0^{(N)}(t,x,c^0):=N\ol{\mg{f}}_0\bigl(t,x/N, c^0\bigr),   \\
&g_0^{(N)}(x,c^0):=N\mg{g}_0\bigl(x/N, c^0\bigr). 
\end{split}
\label{def-scaling}
\ee
Note that, we have
\bea
f_0^{(N)}(t,x,\beta,\vp,\Lambda^0,c^0)=\langle \beta, \vp\rangle+\frac{1}{2}\Blangle \beta,\frac{\Lambda^0}{N}\beta\Brangle
+\ol{f}_0^{(N)}(t,x,c^0).\nn
\eea
Economic meaning of each term can be understood in the same way as the one for minor agents
given in Section~\ref{sec-minor-problem}.

Let us introduce the following assumptions.
\begin{assumption}{\rm (Major)} \\
{\rm (i)} $(\Lambda_t^0)_{t\in[0,T]}$ is an $\ol{\mbb{F}}^0$-progressively measurable $\cals^n$-valued process
such that there exist some positive constants $0<\ul{\lambda}\leq \ol{\lambda}<\infty$  satisfying $\ul{\lambda}|\theta|^2 \leq \blangle \theta, (\Lambda_t^0+2\Lambda_t)\theta\brangle\leq \ol{\lambda}|\theta|^2$
for every $(\omega,t,\theta)\in[0,T]\times \Omega\times \mbb{R}^n$.  \\
{\rm (ii)} For any $(t,c^0)\in [0,T]\times \mbb{R}^n$, 
$|\mg{l}_0(t,c^0)|+|\mg{s}_0(t,c^0)|\leq L_0(1+|c^0|)$.\\
{\rm (iii)} For any $(t,x^0,c^0)\in [0,T]\times (\mbb{R}^n)^2$, 
\be
|\ol{\mg{f}}_0(t,x^0,c^0)|+|\mg{g}_0(x^0,c^0)|\leq L_0(1+|x^0|^2+|c^0|^2). \nn
\ee
{\rm (iv)} $\ol{\mg{f}}_0$ and $\mg{g}_0$ are once continuously differentiable in $x$ and satisfy 
\bea
&&|\part_x \ol{\mg{f}}_0(t,x^0,c^0)|+|\part_x \mg{g}_0(x^0,c^0)|\leq L_0(1+|x^0|+|c^0|), \nn \\
&&|\part_x \ol{\mg{f}}_0(t,x^{0\prime},c^0)-\part_x \ol{\mg{f}}_0(t,x^0,c^0)|
+|\part_x \mg{g}_0(x^{0\prime},c^0)-\part_x \mg{g}_0(x^0,c^0)|\leq L_0|x^{0\prime}-x^0|, \nn
\eea
for any $(t,x^0,x^{0\prime}, c^0)\in [0,T]\times (\mbb{R}^n)^3$. \\
{\rm (v)} $\ol{\mg{f}}_0$ and $\mg{g}_0$ are strictly convex in the sense that
there exist some positive constants $\gamma^f_0, \gamma^g_0>0$ and
\bea
&&\ol{\mg{f}}_0(t,x^{0\prime},c^0)-\ol{\mg{f}}_0(t,x^0,c^0)-\blangle x^{0\prime}-x^0, \part_x \ol{\mg{f}}_0(t,x^0,c^0)\brangle
\geq \frac{\gamma^f_0}{2}|x^{0\prime}-x^0|^2, \nn \\
&&\mg{g}_0(x^{0\prime},c^0)-\mg{g}_0(x^0,c^0)-\blangle x^{0\prime}-x^0, \part_x \mg{g}_0(x^0,c^0)\brangle 
\geq \frac{\gamma^g_0}{2}|x^{0\prime}-x^0|^2, \nn
\eea
hold for any $(t,x^0,x^{0\prime},c^0)\in [0,T]\times (\mbb{R}^n)^3$.
\end{assumption}
For later use, let us put
\be
\gamma_0^{f(N)}:=\frac{\gamma^f_0}{N}, \quad \gamma^{g(N)}_0:=\frac{\gamma^g_0}{N}. \nn
\ee
\begin{remark}
\label{remark-derivative}
With the above definition, we have
\bea
\part_x \ol{f}_0^{(N)}(t,x,c^0)&=&N\frac{\part}{\part x} \ol{\mg{f}}_0(t,x/N, c^0)\nn \\
&=&N\frac{\part (x/N)}{\part x}\part_x\ol{\mg{f}}_0(t,x/N,c^0)=\part_x \ol{\mg{f}}_0(t,x/N,c^0).\nn
\eea
and similar relation for $\part_x \mg{g}_0$.
\end{remark}

\begin{remark}
For the analysis with a fixed $N$, such a scaling is arbitrary and irrelevant. 
However, it plays an important role  when we study the large population limit $N\rightarrow \infty$.
In particular, the market share of the major agent must grow proportionally to the population size $N$.
For example, if the cost functions contain $\blangle \beta, \Lambda^0\beta\brangle$
instead of $\Blangle \beta, \frac{\Lambda^0}{N}\beta\Brangle$, 
the market share of the {\it major} agent becomes negligible in the large population limit.
In this case, we obtain the same market price as in \cite{Fujii-Takahashi, Fujii-mfg-convergence}.
\end{remark}

Following the analysis done in Appendix~\ref{sec-controlled-fbsde}, 
let us introduce the adjoint variables $(p^0, (p^i)_{i=1}^N, (r^i)_{i=1}^N)$ for $(x^0, (x^i)_{i=1}^N, (y^i)_{i=1}^N)$, respectively. 
The (reduced) Hamiltonian 
\bea
\calh:[0,T]\times \mbb{R}^n\times (\mbb{R}^n)^N\times (\mbb{R}^n)^N\times \mbb{R}^n\times (\mbb{R}^n)^N\times (\mbb{R}^n)^N
\times \mbb{R}^n\times \cals^n\times \cals^n_+\times \mbb{R}^n\times (\mbb{R}^n)^N\rightarrow \mbb{R}\nn
\eea
of the system is defined by
\bea
&&\calh(t,x^0, (x^i)_{i=1}^N, (y^i)_{i=1}^N, p^0, (p^i)_{i=1}^N, (r^i)_{i=1}^N, \beta, \Lambda^0,\Lambda, c^0,(c^i)_{i=1}^N)\nn \\
&&\quad:=\blangle p^0,\beta+l_0^{(N)}(t,c^0)\brangle
+\sum_{i=1}^N \Blangle p^i, -\ol{\Lambda}\bigl(y^i-\mg{m}((y))\bigr)-\frac{\beta}{N}+l_i(t,c^0,c^i)\Brangle\nn \\
&&\qquad+\sum_{i=1}^N \blangle r^i, -\part_x \ol{f}_i(t,x^i,c^0,c^i)\brangle\nn \\
&&\qquad+\Blangle \beta, -\mg{m}((y))+\Lambda \frac{\beta}{N}\Brangle+\frac{1}{2}\Blangle \beta,\frac{\Lambda^0}{N}\beta\Brangle
+\ol{f}_0^{(N)}(t,x^0,c^0). 
\label{eq-N-Hamiltonian}
\eea
For a given set of $p^0, (p^i)_{i=1}^N, (r^i)_{i=1}^N$ (and also $(\Lambda^0,\Lambda,c^0,(c^i)_{i=1}^N)$),
it is straightforward to check that $\calh$ is jointly convex in $(x^0, (x^i)_{i=1}^N, (y^i)_{i=1}^N, \beta)$
and strictly convex in $(x^0, \beta)$. Here, recall that $\part_x \ol{f}_i$ is affine in $x^i$ by Assumption (Minor-A, (iv)).
For given inputs, the minimizer of the Hamiltonian $\wh{\beta}:={\rm argmin}\calh(\beta)$ is given by
\bea
\wh{\beta}=N \ol{\calv}^0\bigl(-p^0+\mg{m}((y))+\mg{m}((p))\bigr)
\label{eq-optimal-beta}
\eea
where $\ol{\calv}^0:=(\Lambda^0+2\Lambda)^{-1}$.

The adjoint equations for $(p^0, (p^i)_{i=1}^N, (r^i)_{i=1}^N)$ can be found from $(\ref{A-eq-adjoint})$: for $1\leq i\leq N$,
\bea
\begin{cases}
dP_t^0=-\part_x \ol{f}_0^{(N)}(t,X_t^0,c_t^0)dt+Q_t^{0,0}dW_t^0+\sum_{j=1}^N Q_t^{0,j}dW_t^j, \\
dP_t^i=c_i^f(t,c_t^0,c_t^i)R_t^idt+Q_t^{i,0}dW_t^0+\sum_{j=1}^N Q_t^{i,j}dW_t^j, \\
\display dR_t^i=\Bigl\{\ol{\Lambda}_t\bigl(P_t^i-\mg{m}((P_t))\bigr)+\frac{\beta_t}{N}\Bigr\}dt, 
\end{cases}
\label{eq-adjoint-major}
\eea
with 
\bea
\begin{cases}
P^0_T=\part_x g_0^{(N)}(X_T^0,c_T^0),  \\
P^i_T=-c_i^g(c_T^0,c_T^i)\Bigl(R_T^i+\frac{\del}{1-\del}\mg{m}((R_T))\Bigr),  \\
R_0^i=0 .
\end{cases}
\label{eq-adjoint-terminal-major}
\eea
Note that the forward and backward processes $x$ and $y$ in $(\ref{A-eq-controlled-fbsde})$
correspond to $(X^0, (X^i)_{i=1}^N)$ and $(Y^i)_{i=1}^N$ in $(\ref{eq-state-major})$, respectively.
As for the adjoint processes,  $p$ and $r$ in $(\ref{A-eq-adjoint})$
corresponds to $(P^0, (P^i)_{i=1}^N)$ and $(R^i)_{i=1}^N$, respectively.
By checking Assumption~\ref{assumption-A-1} using the above relations, 
we obtain the next theorem.

\begin{theorem}
\label{th-major-sufficient}
Let Assumptions (Minior-A, B) and (Major) be in force.
Suppose that the system of FBSDEs $(\ref{eq-state-major})$ and $(\ref{eq-adjoint-major})$
with boundary conditions $(\ref{eq-state-terminal-major})$ and $(\ref{eq-adjoint-terminal-major})$
has a solution $X^0, Y^i, P^0, P^i, R^i \in \mbb{S}^2(\mbb{F};\mbb{R}^n)$,
$Z^{i,0}, Q^{0,0}, Q^{i,0}\in \mbb{H}^2(\mbb{F};\mbb{R}^{n\times d_0})$,
and $Z^{i,j}, Q^{0,j}, Q^{i,j}\in \mbb{H}^2(\mbb{F};\mbb{R}^{n\times d})$, $1\leq i,j\leq N$,
with the control process  $\beta_t =\wh{\beta}_t, t\in[0,T)$ i.e.,
\be
\wh{\beta}_t=N \ol{\calv}_t^0\bigl(-P_t^0+\mg{m}((Y_t))+\mg{m}((P_t))\bigr), \quad \ol{\calv}_t^0:=(\Lambda_t^0+2\Lambda_t)^{-1}. \nn
\ee
Then, $(\wh{\beta}_t)_{t\in[0,T)}$ $($with $\wh{\beta}_T=0)$ is the unique optimal control for the major agent.
\begin{proof}
This is the direct result of Theorem~\ref{th-A-verification}. 
Note that Assumption (Minor-A) (iv) plays a crucial role to guarantee the joint convexity of $\calh$
and the affine property of $\Phi$ required in the theorem.
\end{proof}
\end{theorem}

\begin{remark}[on the condition $\beta_T=0$]
In the current work, we restrict the admissible strategies of the major agent to $\{\beta_T=0\}$.
Since $\{t=T\}$ is the Lebesgue null set, $\beta_T$ does not affect the terminal position size $X_T^0$
of the major agent.  Nevertheless,  it affects  the equilibrium price at time $T$  by the relation $(\ref{eq-price-beta})$.
Therefore,  in general,  the major agent has an incentive to manipulate the price  by changing $\beta_T$.
In order to make the optimization problem at $T$ well-defined,  we need a strict convexity in  the terminal cost with respect to $\vp_T$
after including complicated feedback effects from the minor agents. 
Since this makes the analysis intractable for us, we restrict to $\{\beta_T=0\}$ 
and also make $g_0^{(N)}$ independent from $\vp_T$ at the moment. We leave this interesting problem on general $\beta_T$
at the terminal time for future research.
\end{remark}

\subsection{Existence of the optimal solution for the major agent}
From Theorem~\ref{th-major-sufficient}, the crucial target of our analysis is
the following coupled system of FBSDEs:
\bea
\begin{cases}
dX_t^0=\bigl(\wh{\beta}_t+l_0^{(N)}(t,c_t^0)\bigr)dt+\sigma_0^{(N)}(t,c_t^0)dW_t^0,  \\
\display dX_t^i=\Bigl\{-\ol{\Lambda}_t\bigl(Y_t^i-\mg{m}((Y_t))\bigr)-\frac{\wh{\beta}_t}{N}+l_i(t,c^0_t,c_t^i)\Bigr\}dt+
\sigma_i^0(t,c_t^0,c_t^i)dW_t^0+\sigma_i(t,c_t^0,c_t^i)dW_t^i,  \\
\display dR_t^i=\Bigl\{\ol{\Lambda}_t\bigl(P_t^i-\mg{m}((P_t))\bigr)+\frac{\wh{\beta}_t}{N}\Bigr\}dt, \\
dP_t^0=-\part_x \ol{f}_0^{(N)}(t,X_t^0,c_t^0)dt+Q_t^{0,0}dW_t^0+\sum_{j=1}^N Q_t^{0,j}dW_t^j,  \\
dY_t^i=-\part_x \ol{f}_i(t,X_t^i,c_t^0,c_t^i)dt+Z_t^{i,0}dW_t^0+\sum_{j=1}^N Z_t^{i,j}dW_t^j,  \\
dP_t^i=c^f_i(t,c_t^0,c_t^i)R_t^idt+Q_t^{i,0}dW_t^0+\sum_{j=1}^N Q_t^{i,j}dW_t^j, 
\end{cases}
\label{eq-full-coupled}
\eea
with
\bea
\begin{cases}
X_0^0=N\chi^0, ~X_0^i=\xi^i, ~R_0^i=0 \\
P^0_T=\part_x g_0^{(N)}(X_T^0,c_T^0),  \\
Y_T^i=\frac{\del}{1-\del}\mg{m}\Bigl(\bigl(c_j^g(c_T^0,c_T^j)X_T^j+h_j^g(c_T^0,c_T^j)\bigr)_{j=1}^N\Bigr)+c_i^g(c_T^0,c_T^i)X_T^i+h_i^g(c_T^0,c_T^i),  \\
P_T^i=-c_i^g(c_T^0,c_T^i)\Bigl(R_T^i+\frac{\del}{1-\del}\mg{m}((R_T))\Bigr),
\end{cases}
\label{eq-full-terminal}
\eea
for $1\leq i\leq N$. Here,  $(\wh{\beta}_t)_{t\in[0,T)}$ is defined by
\be
\wh{\beta}_t= N \ol{\calv}_t^0\bigl(-P_t^0+\mg{m}((Y_t))+\mg{m}((P_t))\bigr), ~t\in[0,T).  \nn
\ee

The main result of this section is the next theorem.

\begin{theorem}
\label{th-N-major-existence}
Under Assumptions (Minor-A, B) and (Major), there exists a unique strong  solution 
$X^0, Y^i, P^0, P^i, R^i \in \mbb{S}^2(\mbb{F};\mbb{R}^n)$,
$Z^{i,0}, Q^{0,0}, Q^{i,0}\in \mbb{H}^2(\mbb{F};\mbb{R}^{n\times d_0})$,
and $Z^{i,j}, Q^{0,j}, Q^{i,j}\in \mbb{H}^2(\mbb{F};\mbb{R}^{n\times d})$, $1\leq i,j\leq N$
to the coupled system of FBSDEs $(\ref{eq-full-coupled})$ with $(\ref{eq-full-terminal})$.
\begin{proof}
We shall show that the monotone conditions used in Theorem~2.6 in \cite{Peng-Wu} are actually satisfied.
Let $x^0, p^0$ and $x^i, y^i, p^i, r^i, 1\leq i\leq N$ be arbitrary constants in $\mbb{R}^n$.
We put $x=(x^i)_{i=1}^N, y=(y^i)_{i=1}^N, p=(p^i)_{i=1}^N, r=(r^i)_{i=1}^N$, 
and $u=(x^0,x,r,p^0, y,p)$. 
We write $\wh{\beta}(t,u):=\ol{\calv}_t^0\bigl(-p^0+\mg{m}((y))+\mg{m}((p))\bigr)$.
As in Theorem~\ref{th-minor-clearing-existence},
we introduce the quantities:
\be
\begin{split}  
&{\rm drift}[x^0](t,u):=\wh{\beta}(t,u)+l_0^{(N)}(t,c^0_t),\\
&{\rm drift}[x^i](t,u):=-\ol{\Lambda}_t\bigl(y^i-\mg{m}((y))\bigr)-\frac{\wh{\beta}(t,u)}{N}+l_i(t,c_t^0,c_t^i),  \\
&{\rm drift}[r^i](t,u):=\ol{\Lambda}_t\bigl(p^i-\mg{m}((p))\bigr)+\frac{\wh{\beta}(t,u)}{N},   \\
&{\rm drift}[p^0](t,u):=-\part_x \ol{f}_0^{(N)}(t,x^0,c^0_t),  \\
&{\rm drift}[y^i](t,u):=-\part_x \ol{f}_i(t,x^i,c^0_t, c_t^i),  \\
&{\rm drift}[p^i](t,u):=c^f_i(t,c^0_t,c_t^i)r^i, 
\end{split} \nn
\ee
and
\be
\begin{split}
&{\rm terminal}[p^0](u):=\part_x g_0^{(N)}(x^0,c_T^0),   \\
&{\rm terminal}[y^i](u):=\frac{\del}{1-\del}\mg{m}\bigl((c^g_j(c_T^0,c_T^j)x^j+h_j^g(c_T^0,c_T^j))_{j=1}^N\bigr)
+c_i^g(c_T^0,c_T^i)x^i+h_i^g(c_T^0,c_T^i),  \\
&{\rm terminal}[p^i](u):=-c_i^g(c_T^0,c_T^i)\Bigl(r^i+\frac{\del}{1-\del}\mg{m}((r))\Bigr).  
\end{split}\nn
\ee
With two inputs $(u, u^\prime)$, we define $\Del u:=u-u\pp$, 
\bea
&&\Del {\rm drift}[x^0](t):={\rm drift}[x^0](t,u)-{\rm drift}[x^0](t,u^\prime),\nn   \\
&&\Del {\rm terminal}[p^0]:={\rm terminal}[p^0](u)-{\rm temrinal}[p^0](u^\prime), \nn
\eea
and similarly for the others.
From Remark~\ref{remark-derivative}, we have
\bea
\blangle \Del {\rm drift}[p^0](t), \Del x^0\brangle&=&-\blangle \part_x\ol{\mg{f}}_0(t,x^0/N,c_t^0)-\part_x\ol{\mg{f}}_0(t,x^{0\prime}/N,c_t^0),
\Del x_t^0\brangle\nn \\
&\leq& -N\gamma^f_0 |\Del x^0/N|^2=-\gamma_0^{f(N)}|\Del x^0|^2. \nn 
\eea
It is then straightforward to get
\bea
&&\blangle \Del {\rm drift}[p^0](t), \Del x^0\brangle+\sum_{i=1}^N \blangle \Del {\rm drift}[y^i](t), \Del x^i\brangle
+\sum_{i=1}^N \blangle (-I)\Del {\rm drift}[p^i](t), \Del r^i\brangle\nn \\
&&\quad \leq -\gamma^{f(N)}_0|\Del x^0|^2-\gamma^f\sum_{i=1}^N\bigl(|\Del x^i|^2+|\Del r^i|^2\bigr), \nn
\eea
where $I=I_{n\times n}$ is the identity matrix. Next, with $\Del \wh{\beta}_t:=\wh{\beta}(t,u)-\wh{\beta}(t,u\pp)$, 
we have
\be
\begin{split}
&\sum_{i=1}^N \blangle \Del {\rm drift}[x^i](t), \Del y^i\brangle=\sum_{i=1}^N 
\Blangle -\ol{\Lambda}_t\bigl(\Del y^i-\mg{m}((\Del y))\bigr)-\frac{\Del \wh{\beta}_t}{N}, \Del y^i\Brangle  \\
&\quad=-\sum_{i=1}^N \blangle \ol{\Lambda}_t \Del y^i, \Del y^i\brangle+N\blangle \ol{\Lambda}_t \mg{m}((\Del y)), \mg{m}((\Del y))\brangle
-N\Blangle \frac{\Del \wh{\beta}_t}{N}, \mg{m}((\Del y))\Brangle \leq -N\Blangle \frac{\Del \wh{\beta}_t}{N}, \mg{m}((\Del y))\Brangle. 
\end{split}\nn
\ee
By similar calculation, we get
\bea
&&\blangle \Del {\rm drift}[x^0](t), \Del p^0\brangle+\sum_{i=1}^N \blangle \Del {\rm drift}[x^i](t), \Del y^i\brangle
+\sum_{i=1}^N \blangle (-I)\Del{\rm drift}[r^i](t), \Del p^i\brangle \nn \\
&&\quad\leq -N\Blangle \frac{\Del \wh{\beta}_t}{N}, -\Del p^0+\mg{m}((\Del y))+\mg{m}((\Del p))\Brangle
=-N\Blangle \frac{\Del \wh{\beta}_t}{N}, (\Lambda_t^0+2\Lambda_t)\frac{\Del \wh{\beta}_t}{N}\Brangle \leq 0.\nn
\eea
Therefore, from the drift contribution, we eventually have 
\bea
&&\blangle \Del {\rm drift}[p^0](t), \Del x^0\brangle+\sum_{i=1}^N \blangle \Del {\rm drift}[y^i](t), \Del x^i\brangle
+\sum_{i=1}^N \blangle (-I)\Del {\rm drift}[p^i](t), \Del r^i\brangle\nn \\
&&+\blangle \Del {\rm drift}[x^0](t), \Del p^0\brangle+\sum_{i=1}^N \blangle \Del {\rm drift}[x^i](t), \Del y^i\brangle
+\sum_{i=1}^N \blangle (-I)\Del{\rm drift}[r^i](t), \Del p^i\brangle \nn \\
&&\qquad \leq -\gamma^{f(N)}_0|\Del x^0|^2-\gamma^f\sum_{i=1}^N\bigl(|\Del x^i|^2+|\Del r^i|^2\bigr).
\label{eq-N-mono-drift}
\eea

For the terminal conditions, by the similar calculation done in $(\ref{eq-terminal-cal})$, we obtain
\bea
&&\blangle \Del {\rm terminal}[p^0],\Del x^0\brangle+\sum_{i=1}^N \blangle \Del {\rm terminal}[y^i], \Del x^i\brangle
+\sum_{i=1}^N \blangle (-I) \Del {\rm terminal}[p^i], \Del r^i\brangle\nn \\
&&\quad \geq \gamma^{g(N)}_0|\Del x^0|^2+(\gamma^g-\mg{a})\sum_{i=1}^N \bigl(|\Del x^i|^2+|\Del r^i|^2\bigr). 
\label{eq-N-mono-terminal}
\eea
Using $(\ref{eq-N-mono-drift})$ and $(\ref{eq-N-mono-terminal})$, we can now apply Theorem~2.6 in \cite{Peng-Wu} with
\bea
&&A(t,u)=\begin{pmatrix}
I_{n\times n}  & 0 & 0& 0 & 0 & 0\\
0 & (I_{n\times n})^N & 0& 0& 0 & 0 \\
0 & 0 & (-I_{n\times n})^N & 0 & 0& 0 \\
0 & 0 & 0 & I_{n\times n} & 0& 0 \\
0 & 0 & 0 & 0 & (I_{n\times n})^N& 0 \\
0 & 0 & 0 & 0 & 0& (-I_{n\times n})^N 
\end{pmatrix}
\begin{pmatrix}
{\rm drift}[p^0]\\
{\rm drift}[y]\\
{\rm drift}[p]\\
{\rm drift}[x^0]\\
{\rm drift}[x]\\
{\rm drift}[r]
\end{pmatrix} (t,u)\nn
\eea
and
\bea
G=\begin{pmatrix}
I_{n\times n}  & 0 & 0 \\
0 & (I_{n\times n})^N & 0 \\
0 & 0 & (-I_{n\times n})^N  \\
\end{pmatrix}. \nn
\eea
In particular, we have $\beta_1:=\min(\gamma^{f(N)}_0, \gamma^f)>0$,  $\mu_1:=\min(\gamma^{g(N)}_0, \gamma^g-\mg{a})>0$.
Note that the coefficients of the Brownian motions ($\sigma_i$ etc.) are irrelevant since they are uncontrolled and state-independent.
In fact, one can repeat the proof for Theorem~3.3 in \cite{Fujii-mfg-convergence} 
in essentially the same way by simply replacing 
the analysis for $d\blangle \Del y_t, \Del x_t\brangle$ with
that for
\bea
d\left\langle \begin{pmatrix} \Del p^0_t \\ \Del y_t \\ \Del p_t\end{pmatrix},
G\begin{pmatrix} \Del x^0_t \\ \Del x_t \\ \Del r_t\end{pmatrix} \right\rangle \nn
\eea
using the above estimates.
\end{proof}
\end{theorem}

Thanks to Theorem~\ref{th-N-major-existence}, 
we now find the market-clearing price process is given by
\bea
\vp_t=-\mg{m}((Y_t))+\Lambda_t \ol{\calv}_t^0 \Bigl(-P_t^0+\mg{m}((Y_t))+\mg{m}((P_t))\Bigr), \quad t\in[0,T)
\label{eq-price-hetero}
\eea
using the solutions to the system of FBSDEs $(\ref{eq-full-coupled})$ with $(\ref{eq-full-terminal})$.
Note that the system of equations is {\it coupled} among the agents $1\leq i \leq N$ 
by the interactions through the empirical means such as $\mg{m}((Y_t))$.
As in the case for the standard mean field game theory for Nash equilibrium,
we can obtain a simpler {\it decoupled} system described by the FBSDE of McKean-Vlasov type
in the large-$N$ limit. This is the major topic to be treated in the remainder of the work. 
In the next section, we study the mean-field limit of the corresponding problem under the assumption that the minor agents are homogeneous.

\begin{remark}[on Nash equilibrium]
The equilibrium in our model is characterized by the relation
$\sum_{i=1}^N \wh{\alpha}_t^i+\wh{\beta}_t=0$, $dt\otimes d\mbb{P}$-a.e. 
Notice that this market-clearing equilibrium is  a different concept from  the Nash equilibrium.
Since Nash equilibrium is characterized by the optimality of the value function of each agent with respect to his/her strategy while keeping the other agents' strategies unchanged, it inevitably violates the market-clearing condition and hence is inapplicable to our case.
In fact, because of this reason, the market-clearing equilibrium is quite popular in standard economic theories.
\end{remark}

\section{Mean-field Equilibrium}

Let us work on the probability space with $N=1$ in
Section~\ref{sec-notation},   i.e.  $(\Omega,\calf, \mbb{P}, \mbb{F})=(\Omega^1, \calf^1, \mbb{P}^1, \mbb{F}^1)$.
In the following, we use the notation:
\be
\mbb{E}_t^0\bigl[\cdot\bigr]:=\mbb{E}\bigl[\cdot~|\ol{\calf}_t^0\bigr]. \nn
\ee

Let us first introduce the following assumptions.
\begin{assumption}{\rm (MFG)}\\
{\rm (i)} $(l,\sigma^0, \sigma, \ol{f}, \ol{g},  c^f, c^g, h^f, h^g)$ satisfy the same conditions
corresponding to those for  $(l_i,\sigma_i^0, \sigma_i, \ol{f}_i, \ol{g}_i,$  $~~c_i^f, c_i^g, h_i^f, h_i^g)$ in Assumption (Minor-A). \\
{\rm (ii)} There exists some $\ol{\calf}^0_T$-measurable $\cals^n$-valued random variable $\mg{c}$ such that
\bea
\mg{a}:=\frac{\del}{1-\del}||\mg{c}-c^g(c^0_T, c_T^1)||_{\infty}<\gamma^g. \nn 
\eea
{\rm (iii)} For the other variables and functions, we assume the same conditions as those in Assumptions (Minor-A) and (Major).
\end{assumption}

For the space of admissible strategies $\mbb{A}^0_{\rm mfg}:=\mbb{H}^2(\ol{\mbb{F}}^0;\mbb{R}^n)\cap\{\beta_T=0\}$,
we suppose that the major agent tries to  solve 
\be
\inf_{\bg{\beta}\in \mbb{A}^{0}_{\rm mfg}}\calj_0(\bg{\beta})
\label{problem-mfg-major}
\ee
where
\bea
\calj_0(\bg{\beta}):=\mbb{E}\Bigl[\int_0^T \mg{f}_0\Bigl(t,x_t^0,\beta_t, -\mbb{E}_t^0[y_t^1]+\Lambda_t \beta_t, 
\Lambda^0_t, c_t^0\Bigr)dt+\mg{g}_0(x_T^0, c_T^0)\Bigr]\nn
\eea
subject to the following dynamic constraints:
\bea
\begin{cases}
dx_t^0=\bigl(\beta_t+\mg{l}_0(t,c_t^0)\bigr)dt+\mg{s}_0(t,c_t^0)dW_t^0,  \\
dx_t^1=\Bigl\{-\ol{\Lambda}_t \bigl(y_t^1-\mbb{E}_t^0[y_t^1]\bigr)-\beta_t+l(t,c_t^0,c_t^1)\Bigr\}dt+\sigma^0(t,c_t^0,c_t^1)dW_t^0
+\sigma(t,c_t^0,c_t^1)dW_t^1,  \\
dy_t^1=-\part_x \ol{f}(t,x_t^1, c_t^0,c_t^1)dt+z_t^{1,0}dW_t^0+z_t^{1,1}dW_t^1, 
\end{cases}
\label{major-control-mfg}
\eea
with 
\bea
\begin{cases}
x_0^0=\chi^0, \quad x_0^1=\xi^1, \\
\display y_T^1=\frac{\del}{1-\del}\mbb{E}_T^0\bigl[c^g(c_T^0,c_T^1)x_T^1+h^g(c_T^0,c_T^1)\bigr]+c^g(c_T^0,c_T^1)x_T^1+h^g(c_T^0,c_T^1).
\end{cases}
\label{major-control-mfg-terminal}
\eea
Here, the problem for the major agent is the optimization with respect to the controlled-FBSDE
of conditional McKean-Vlasov type.  One can naturally expect the above formulation of the problem
in the mean-field limit from the McKean-Vlasov FBSDEs given in \cite{Fujii-Takahashi}
and the expression in $(\ref{eq-price-beta})$

\begin{remark}
\label{remark-mfg-well-posed}
Notice that, the above problem is well posed in the sense that
for a given $\bg{\beta}\in \mbb{A}^0_{\rm mfg}$, there exists a unique strong solution to $(\ref{major-control-mfg})$
and the corresponding cost $\calj_0(\bg{\beta})$ is finite.
In particular, the unique existence for $(x^1, y^1)$ can be proved
by a simple modification of Theorem~4.2 in \cite{Fujii-Takahashi}.
\end{remark}

Implied from $(\ref{eq-N-Hamiltonian})$, we consider the Hamiltonian 
\be
H:[0,T]\times (\mbb{R}^n)^9\times \cals^n\times \cals^n_+\times (\mbb{R}^n)^2\rightarrow \mbb{R} \nn
\ee
by
\bea
&&H\bigl(t,x^0, x^1, y^1, \ol{y}^1, p^0, p^1, \ol{p}^1, r^1, \beta, \Lambda^0, \Lambda, c^0, c^1\bigr) \nn \\
&&\quad:=\blangle p^0,\beta+\mg{l}_0(t,c^0)\brangle+\blangle p^1, -\ol{\Lambda}(y^1-\ol{y}^1)+l(t,c^0,c^1)\brangle
+\blangle \ol{p}^1, -\beta\brangle \nn \\
&&\quad+\blangle r^1, -\part_x \ol{f}(t,x^1,c^0,c^1)\brangle+\blangle \beta, -\ol{y}^1+\Lambda\beta\brangle
+\frac{1}{2}\blangle \beta, \Lambda^0\beta\brangle+\ol{\mg{f}}_0(t,x^0,c^0).
\label{eq-mfg-Hamiltonian}
\eea
It is important to observe that the map 
\bea
(x^0, x^1, y^1, \ol{y}^1, \beta)\mapsto H(t,x^0,x^1, y^1, \ol{y}^1, p^0, p^1, \ol{p}^1, r^1, \beta, \Lambda^0, \Lambda, c^0,c^1) \nn
\eea
is jointly convex and strictly convex in $\beta$ (and $x^0$). It is easy to find
\bea
\wh{\beta}=\ol{\calv}^0\bigl(-p^0+\ol{y}^1+\ol{p}^1\bigr), \nn
\eea
with $\ol{\calv}^0:=(\Lambda^0+2\Lambda)^{-1}$ gives the minimizer of $H$ with respect to $\beta$.

The relevant set of adjoint equations can be inferred from Appendix~\ref{sec-controlled-fbsde}
combined with Chapter 6 in \cite{Carmona-Delarue-1}, or from $(\ref{eq-full-coupled})$ and $(\ref{eq-full-terminal})$.
\bea
\begin{cases}
dr_t^1=\Bigl\{\ol{\Lambda}_t\bigl(p_t^1-\mbb{E}_t^0[p_t^1]\bigr)+\beta_t\Bigr\}dt,  \\
dp_t^0=-\part_x \ol{\mg{f}}_0(t,x_t^0,c_t^0)dt+q_t^{0,0}dW_t^0,  \\
dp_t^1=c^f(t,c_t^0,c_t^1)r_t^1dt+q_t^{1,0}dW^0_t+q_t^{1,1}dW_t^1, 
\end{cases}
\label{eq-adjoint-mfg}
\eea
with
\bea
\begin{cases}
r_0^1=0, \\
p_T^0=\part_x \mg{g}_0(x_T^0,c_T^0), \\
p_T^1=-c^g(c_T^0,c_T^1)\Bigl(r_T^1+\frac{\del}{1-\del}\mbb{E}^0_T[r_T^1]\Bigr). 
\end{cases}
\label{eq-adjoint-mfg-terminal}
\eea

In order to guarantee that the above system characterizes the optimal solution for the major agent,
we are going to prove the new verification theorem for controlled FBSDEs of conditional McKean-Vlasov type. 
\begin{theorem}
\label{th-mfg-verification}
Let Assumption (MFG) be in force. Suppose 
that there exists a solution $(\wh{x}^0, \wh{x}^1, \wh{r}^1, \wh{p}^0, \wh{p}^1, \wh{y}^1)$ to 
$((\ref{major-control-mfg}), (\ref{major-control-mfg-terminal}))$
and  $((\ref{eq-adjoint-mfg}), (\ref{eq-adjoint-mfg-terminal}))$ with the control process $\beta$ satisfying
\be
\wh{\beta_t}=\ol{\calv}_t^0\bigl(-\wh{p}_t^0+\mbb{E}_t^0[\wh{y}_t^1]+\mbb{E}_t^0[\wh{p}_t^1]\bigr)\nn 
\ee
$dt\otimes d\mbb{P}$-a.e., then $(\wh{\beta}_t)_{t\in[0,T)}$ $($with $\wh{\beta}_T=0)$ is the unique optimal control 
for the problem $(\ref{problem-mfg-major})$.
\begin{proof}
For a given $\beta\in \mbb{A}^0_{\rm mfg}$, we denote the associated solution to $(\ref{major-control-mfg})$
by $(x^0,x^1, y^1)$.
We shall study the difference:
\bea
&&\calj_0(\bg{\beta})-\calj_0(\wh{\bg{\beta}})=\mbb{E}\Bigl[\mg{g}_0(x_T^0, c_T^0)-\mg{g}_0(\wh{x}_T^0, c_T^0) \nn \\
&&\quad+\int_0^T \Bigl(\mg{f}_0\bigl(t,x_t^0,\beta_t,-\mbb{E}_t^0[y^1_t]+\Lambda_t \beta_t, \Lambda_t^0, c_t^0\bigr)
-\mg{f}_0\bigl(t,\wh{x}_t^0,\wh{\beta}_t,-\mbb{E}_t^0[\wh{y}_t^1]+\Lambda_t \wh{\beta}_t, \Lambda_t^0, c_t^0\bigr)\Bigr)dt\Bigr].\nn
\eea
First, observe that
\bea
&&\mbb{E}\Bigl[\blangle \wh{p}_T^1, x_T^1-\wh{x}_T^1\brangle+\blangle \wh{r}_T^1, y_T^1-\wh{y}_T^1\brangle\Bigr]\nn \\
&&\quad=\frac{\del}{1-\del}\mbb{E}\Bigl[-\blangle \mbb{E}^0_T[\wh{r}_T^1], c^g(c_T^0,c_T^1)(x_T^1-\wh{x}_T^1)\brangle
+\blangle \wh{r}_T^1, \mbb{E}_T^0[c^g(c_T^0,c_T^1)(x_T^1-\wh{x}_T^1)]\brangle\Bigr]\nn \\
&&\quad =0.\nn
\eea
Thus, from the convexity of $\mg{g}_0$, we have
\bea
&&\mbb{E}\Bigl[\mg{g}_0(x_T^0, c_T^0)-\mg{g}_0(\wh{x}_T^0, c_T^0)\Bigr] \geq \mbb{E}\Bigl[\blangle \wh{p}_T^0, x_T^0-\wh{x}_T^0\brangle+\blangle \wh{p}_T^1, x_T^1-\wh{x}_T^1\brangle
+\blangle \wh{r}_T^1, y_T^1-\wh{y}_T^1\brangle \Bigr]. 
\label{eq-mfg-terminal-convexity}
\eea
Let us use $\wh{\Theta}_t:=\bigl(\wh{x}_t^0, \wh{x}_t^1, \wh{y}^1, \mbb{E}_t^0[\wh{y}_t^1], \wh{p}_t^0, \wh{p}_t^1, \mbb{E}_t^0[\wh{p}_t^1],
\wh{r}^1_t\bigr)$, $\wh{\theta}_t:=\bigl(\wh{p}_t^0, \wh{p}_t^1, \mbb{E}_t^0[\wh{p}_t^1], \wh{r}_t^1\bigr)$
and omit the common arguments $(\Lambda^0_t, \Lambda_t,c^0_t, c_t^1)$ in the Hamiltonian.
Since $\ol{\Lambda}, \wh{\beta}$ are $\ol{\mbb{F}}^0$-adapted, we  have
\be
\begin{split}
&\mbb{E}\bigl[\bigl\langle \ol{\Lambda}_t (\wh{p}^1_t-\mbb{E}_t^0[\wh{p}^1_t])+\wh{\beta}_t, y_t^1-\wh{y}^1_t\bigr\rangle \bigr] \\
&\quad=\mbb{E}\bigl[\bigl\langle \ol{\Lambda}_t \wh{p}_t^1, y^1_t-\wh{y}^1_t\bigr\rangle \bigr]
+\mbb{E}\bigl[\bigl\langle -\ol{\Lambda}_t \mbb{E}_t^0[\wh{p}^1_t]+\wh{\beta}_t, y_t^1-\wh{y}^1_t\bigr\rangle \bigr] \\
&\quad =\mbb{E}\bigl[\bigl\langle \ol{\Lambda}_t \wh{p}^1_t, y^1_t-\wh{y}^1_t\bigr\rangle \bigr]
+\mbb{E}\bigl[\bigl\langle -\ol{\Lambda}_t \wh{p}^1_t+\wh{\beta}_t, \mbb{E}_t^0[y_t^1]-\mbb{E}_t^0[\wh{y}^1_t]\bigr\rangle \bigr] \\
&\quad =\mbb{E}\bigl[-\bigl\langle \part_{y^1} H(t,\wh{\Theta}_t,\wh{\beta}_t), y_t^1-\wh{y}^1_t\bigr\rangle
-\bigl\langle \part_{\ol{y}^1}H(t,\wh{\Theta}_t,\wh{\beta}_t), \mbb{E}_t^0[y_t^1]-\mbb{E}_t^0[\wh{y}_t^1]\bigr\rangle \bigr].
\end{split} \nn
\ee

With these results and $(\ref{eq-mfg-terminal-convexity})$, a simple application of \Ito-formula yields
\be
\begin{split}
&\calj_0(\bg{\beta})-\calj_0(\wh{\bg{\beta}}) \\
&\geq\mbb{E}\int_0^T \Bigl[
H\bigl(t,x^0_t,x_t^1,y_t^1,\mbb{E}_t^0[y_t^1],\wh{\theta}_t,\beta_t\bigr)
-H\bigl(t,\wh{\Theta}_t, \wh{\beta}_t\bigr) \\
&\quad\qquad -\bigl\langle \part_{x^0}H(t,\wh{\Theta}_t,\wh{\beta}_t), x^0_t-\wh{x}_t^0\bigr\rangle
-\bigl\langle \part_{x^1} H(t, \wh{\Theta}_t,\wh{\beta}_t), x_t^1-\wh{x}^1_t\bigr\rangle \\
&\quad\qquad-\bigl\langle \part_{y^1} H(t,\wh{\Theta}_t,\wh{\beta}_t), y_t^1-\wh{y}^1_t\bigr\rangle
-\bigl\langle \part_{\ol{y}^1}H(t,\wh{\Theta}_t,\wh{\beta}_t), \mbb{E}_t^0[y_t^1]-\mbb{E}_t^0[\wh{y}^1_t]\bigr\rangle \Bigr]dt \\
\end{split}\nn
\ee
\be
\begin{split}
&\geq\mbb{E}\int_0^T \Bigl[
H\bigl(t,x^0_t,x_t^1,y_t^1,\mbb{E}_t^0[y_t^1],\wh{\theta}_t,\beta_t\bigr)
-H\bigl(t,\wh{\Theta}_t, \wh{\beta}_t\bigr)\nn \\
&\quad\qquad -\bigl\langle \part_{x^0}H(t,\wh{\Theta}_t,\wh{\beta}_t), x^0_t-\wh{x}_t^0\bigr\rangle
-\bigl\langle \part_{x^1} H(t, \wh{\Theta}_t,\wh{\beta}_t), x_t^1-\wh{x}^1_t\bigr\rangle\nn \\
&\quad\qquad-\bigl\langle \part_{y^1} H(t,\wh{\Theta}_t,\wh{\beta}_t), y_t^1-\wh{y}^1_t\bigr\rangle
-\bigl\langle \part_{\ol{y}^1}H(t,\wh{\Theta}_t,\wh{\beta}_t), \mbb{E}_t^0[y_t^1]-\mbb{E}_t^0[\wh{y}^1_t]\bigr\rangle  \nn \\
&\quad\qquad -\blangle \part_\beta H(t,\wh{\Theta}_t,\wh{\beta}_t), \beta_t-\wh{\beta}_t\brangle
\Bigr]dt\nn \\
&\geq 0, 
\end{split}\nn
\ee
where the second inequality follows from the fact that $\wh{\beta}_t={\rm argmin}_{\beta}H(t,\wh{\Theta}_t,\beta)$.
The equality holds only when $\beta=\wh{\beta}$ due to the strict convexity.
\end{proof}
\end{theorem}

From Theorem~\ref{th-mfg-verification}, it is clear that the relevant set of equations is given by
\bea
\begin{cases}
dx_t^0=\bigl(\wh{\beta}_t+\mg{l}_0(t,c_t^0)\bigr)dt+\mg{s}_0(t,c_t^0)dW_t^0,  \\
dx_t^1=\Bigl\{-\ol{\Lambda}_t\bigl(y_t^1-\mbb{E}_t^0[y_t^1]\bigr)-\wh{\beta}_t+l(t,c_t^0,c_t^1)\Bigr\}dt
+\sigma^0(t,c_t^0,c_t^1)dW_t^0+\sigma(t,c_t^0,c_t^1)dW_t^1,  \\
dr_t^1=\Bigl\{\ol{\Lambda}_t\bigl(p_t^1-\mbb{E}_t^0[p_t^1]\bigr)+\wh{\beta}_t\Bigr\}dt,  \\
dp_t^0=-\part_x \ol{\mg{f}}_0(t,x_t^0,c_t^0)dt+q_t^{0,0}dW_t^0,  \\
dy_t^1=-\part_x \ol{f}(t,x_t^1,c_t^0,c_t^1)dt+z_t^{1,0}dW_t^0+z_t^{1,1}dW_t^1,  \\
dp_t^1=c^f(t,c_t^0,c_t^1)r_t^1dt+q_t^{1,0}dW_t^0+q_t^{1,1}dW_t^1, 
\end{cases}
\label{eq-mfg-full}
\eea
with
\bea
\begin{cases}
x^0_0=\chi^0, \quad x_0^1=\xi^1, \quad r_0^1=0,   \\
p_T^0=\part_x \mg{g}_0(x_T^0,c_T^0),  \\
\display y_T^1=\frac{\del}{1-\del}\mbb{E}_T^0\bigl[c^g(c_T^0,c_T^1)x_T^1+h^g(c_T^0,c_T^1)\bigr]+c^g(c_T^0,c_T^1)x_T^1+h^g(c_T^0,c_T^1),  \\
\display p_T^1=-c^g(c_T^0,c_T^1)\Bigl(r_T^1+\frac{\del}{1-\del}\mbb{E}_T^0[r_T^1]\Bigr),
\end{cases}
\label{eq-mfg-full-terminal}
\eea
where $\wh{\beta}_t,~t\in[0,T)$ is defined by
\bea
\wh{\beta}_t:=\ol{\calv}_t^0\bigl(-p_t^0+\mbb{E}^0_t[y^1_t]+\mbb{E}^0_t[p_t^1]\bigr). \nn
\eea
The next theorem guarantees the existence of the solution to the above FBSDE and hence the 
optimal control for the major agent in the mean-field limit.

\begin{theorem}
\label{th-mfg-full-existence}
Under the Assumption (MFG), there exists a unique strong solution 
$x^0, p^0\in \mbb{S}^2(\ol{\mbb{F}}^0;\mbb{R}^n)$, $x^1, r^1,  y^1, p^1\in \mbb{S}^2(\mbb{F}^1;\mbb{R}^n)$, 
$q^{0,0}\in \mbb{H}^2(\ol{\mbb{F}}^0;\mbb{R}^{n\times d_0})$, $z^{1,0}, q^{1,0}\in \mbb{H}^2(\mbb{F}^1;\mbb{R}^{n\times d_0})$
and $z^{1,1}, q^{1,1}\in \mbb{H}^2(\mbb{F}^1;\mbb{R}^{n\times d})$ 
to the system of FBSDEs of conditional
McKean-Vlasov type $(\ref{eq-mfg-full})$ with $(\ref{eq-mfg-full-terminal})$. 
\begin{proof}
As we have done in the proof for Theorem~\ref{th-N-major-existence}, we introduce  
$u:=(x^0, x^1, r^1, p^0, y^1, p^1)$ as arbitrary square integrable random variables with 
appropriate dimensions satisfying that $(x^0, p^0)$ are $\ol{\calf}_t^0$-measurable, and the others 
are $\calf^1_t$-measurable. For these inputs, we define
\begin{equation}
\begin{split}
&{\rm drift}[x^0](t,u):=\wh{\beta}(t,u)+\mg{l}_0(t,c_t^0),  \\
&{\rm drift}[x^1](t,u):=-\ol{\Lambda}_t(y^1-\mbb{E}_t^0[y^1])-\wh{\beta}(t,u)+l(t,c_t^0,c_t^1), \\
&{\rm drift}[r^1](t,u):=\ol{\Lambda}_t(p^1-\mbb{E}^0_t[p^1])+\wh{\beta}(t,u),  \\
&{\rm drift}[p^0](t,u):=-\part_x \ol{\mg{f}}_0(t,x^0,c^0_t), \\
&{\rm drift}[y^1](t,u):=-\part_x\ol{f}(t,x^1, c^0_t,c_t^1),  \\
&{\rm drift}[p^1](t,u):=c^f(t,c_t^0,c_t^1)r^1,  
\end{split}\nn
\end{equation}
where $\wh{\beta}(t,u):=\ol{\calv}_t^0(-p^0+\mbb{E}^0_t[y^1]+\mbb{E}^0_t[p^1])$.
For two different inputs $u, u\pp$, we set $\Del u:=u-u\pp$, 
$\Del {\rm drift}[x^0](t):={\rm drift}[x^0](t,u)-{\rm drift}[x^0](t,u\pp)$
and similarly for the other quantities, too.
Since $\ol{\calv}_t^0$ and $p^0$ are $\ol{\calf}_t^0$-measurable, we see
\be
\begin{split}
&\mbb{E}_t^0\bigl[\blangle \Del {\rm drift}[x^0](t), \Del p^0\brangle+
\blangle \Del {\rm drift}[x^1](t), \Del y^1\brangle+\blangle (-I)\Del {\rm drift}[r^1](t), \Del p^1\brangle\bigr] \\
&\quad=-\mbb{E}^0_t\bigl[ \blangle \ol{\calv}_t^0(-\Del p^0+\mbb{E}^0_t[\Del y^1+\Del p^1]),
-\Del p^0+\Del y^1+\Del p^1\brangle\bigr] \\
&\quad=-\blangle \ol{\calv}_t^0(-\Del p^0+\mbb{E}^0_t[\Del y^1+\Del p^1]), -\Del p^0+
\mbb{E}_t^0[\Del y^1+\Del p^1]\brangle\leq 0. 
\end{split}
\label{mfg-monotone-drift}
\ee
It is straightforward to check
\be
\begin{split}
&\mbb{E}\bigl[ \blangle \Del {\rm drift}[p^0](t), \Del x^0\brangle+
\blangle \Del {\rm drift}[y^1](t), \Del x^1\brangle+\blangle (-I)\Del {\rm drift}[p^1](t), \Del r^1\brangle \bigr] \\
&~\leq -\gamma^f_0\mbb{E}|\Del x^0|^2-\gamma^f\bigl(\mbb{E}|\Del x^1|^2+\mbb{E}|\Del r^1|^2\bigr). 
\end{split}
\label{mfg-monotone-driver}
\ee

Now we set $v:=(x^0, x^1, r^1)$ as arbitrary square integrable random variables with 
appropriate dimensions satisfying that $x^0$ are $\ol{\calf}_T^0$-measurable, and the others 
are $\calf^1_T$-measurable. For these inputs, let us define
\be
\begin{split}
&{\rm terminal}[p^0](v):=\part_x\mg{g}_0(x^0, c_T^0),  \\
&{\rm terminal}[y^1](v):=\frac{\del}{1-\del}\mbb{E}_T^0\bigl[c^g(c_T^0,c_T^1)x^1+h^g(c_T^0,c_T^1)\bigr]+c^g(c_T^0,c_T^1)x^1+h^g(c_T^0,c_T^1), \\
&{\rm terminal}[p^1](v):=-c^g(c_T^0,c_T^1)\Bigl(r^1+\frac{\del}{1-\del}\mbb{E}^0_T[r^1]\Bigr),
\end{split} \nn
\ee
and with two different input $v, v\pp$, we denote by $\Del v:=v-v\pp$,
\be
\Del {\rm terminal}[p^0]:={\rm terminal}[p^0](v)-{\rm terminal}[p^0](v\pp)\nn
\ee
and similarly for the other quantities. Observe that
\be
\begin{split}
&\mbb{E}^0_T\bigl[\blangle \Del {\rm terminal}[y^1], \Del x^1\brangle\bigr] \\
&\quad=\mbb{E}^0_T\Bigl[ \Blangle \frac{\del}{1-\del}\mbb{E}^0_T[c^g(c_T^0,c_T^1)\Del x^1]+c^g(c_T^0,c_T^1)\Del x^1, \Del x^1\Brangle\Bigr]\\
&\quad \geq \gamma^g \mbb{E}^0_T |\Del x^1|^2+\frac{\del}{1-\del}\blangle \mg{c}\mbb{E}^0_T[\Del x^1], \mbb{E}^0_T[\Del x^1]\brangle
+\frac{\del}{1-\del}\blangle \mbb{E}^0_T[(c^g(c^0_T, c^1_T)-\mg{c})\Del x^1], \mbb{E}^0_T[\Del x^1]\brangle \\
&\quad \geq (\gamma^g-\mg{a})\mbb{E}^0_T|\Del x^1|^2. 
\end{split}\nn
\ee
Similar calculation yields
\be
\begin{split}
&\mbb{E}\bigl[\blangle \Del {\rm terminal}[p^0], \Del x^0\brangle+
\blangle \Del {\rm terminal}[y^1], \Del x^1\brangle+
\blangle (-I)\Del {\rm terminal}[p^1], \Del r^1\brangle\bigr]\\
&\quad \geq \gamma^g_0\mbb{E}|\Del x^0|^2+(\gamma^g-\mg{a})\mbb{E}\bigl[|\Del x^1|^2+|\Del r^1|^2\bigr].
\end{split}
\label{mfg-monotone-terminal}
\ee

We have now  obtained the monotone conditions necessary for the method of continuation.
In particular, by introducing a strictly positive constant $\gamma>0$
\be
\gamma:=\min\bigl\{\gamma^f_0, \gamma^f, \gamma^g_0, \gamma^g-\mg{a}\bigr\}, \nn
\ee 
we have from $(\ref{mfg-monotone-drift})$, $(\ref{mfg-monotone-driver})$ and $(\ref{mfg-monotone-terminal})$
\be
\begin{split}
&\mbb{E}\bigl[\blangle \Del {\rm drift}[x^0](t), \Del p^0\brangle+
\blangle \Del {\rm drift}[x^1](t), \Del y^1\brangle+\blangle (-I)\Del {\rm drift}[r^1](t), \Del p^1\brangle\bigr] \leq 0, \\
&\mbb{E}\bigl[ \blangle \Del {\rm drift}[p^0](t), \Del x^0\brangle+
\blangle \Del {\rm drift}[y^1](t), \Del x^1\brangle+\blangle (-I)\Del {\rm drift}[p^1](t), \Del r^1\brangle \bigr] \\
&\quad \leq -\gamma \mbb{E}\bigl[|\Del x^0|^2+|\Del x^1|^2+|\Del r^1|^2\bigr], \\
&\mbb{E}\bigl[\blangle \Del {\rm terminal}[p^0], \Del x^0\brangle+
\blangle \Del {\rm terminal}[y^1], \Del x^1\brangle+
\blangle (-I)\Del {\rm terminal}[p^1], \Del r^1\brangle\bigr]\\
&\quad \geq \gamma \mbb{E}\bigl[|\Del x^0|^2+|\Del x^1|^2+|\Del r^1|^2\bigr].
\end{split} 
\label{mfg-mono-summary}
\ee

We can now repeat  the proof of \cite[Theorem 4.2]{Fujii-Takahashi}.
The three conditions of $(\ref{mfg-mono-summary})$ correspond to those of $(4.3)$ (with $L_\vp=0$) in \cite{Fujii-Takahashi}. 
We treat $(x^0_t, x^1_t,r^1_t)$ and $(p^0_t, y_t^1, p_t^1)$ as the tuple of forward 
and backward processes, which are represented by $X_t$ and $Y_t$ in \cite{Fujii-Takahashi}, respectively.
By replacing  the estimate on $\mbb{E}[\blangle \Del Y_T, \Del X_T\brangle]$ 
by
\bea
\mbb{E}\left[\left\langle \begin{pmatrix} \Del p^0_T \\ \Del y_T^1 \\ \Del p_T^1\end{pmatrix},
G\begin{pmatrix} \Del x^0_T \\ \Del x_T^1 \\ \Del r_T^1\end{pmatrix} \right\rangle \right]\nn
\eea
with
$G=\begin{pmatrix}
I_{n\times n}  & 0 & 0 \\
0 & (I_{n\times n}) & 0 \\
0 & 0 & (-I_{n\times n})  \\
\end{pmatrix}$,
we can follow the same procedures with slightly more cumbersome indexing.
\end{proof}
\end{theorem}

\begin{remark}[on Lasry-Lions monotonicity]
The so-called Lasry-Lions monotonicity is a famous criterion for the uniqueness of the mean field games.
It dates back to their original papers~\cite{Lions-1, Lions-2, Lions-3}
and is defined as follows~\cite[Definition 3.28]{Carmona-Delarue-1}:
a real-valued function $h$ on $\mbb{R}^d\times \calp_2(\mbb{R}^d)$ is 
said to be monotone in the sense of Lasry and Lions, if, for all $\mu\in\calp_2(\mbb{R}^d)$,
the mapping $\mbb{R}^d\ni x\mapsto h(x,\mu)$ is at most quadratic 
growth,  and for all $\mu, \mu^\prime \in \calp_2(\mbb{R}^d)$, we have
\be
\int_{\mbb{R}^d}\bigl(h(x,\mu)-h(x,\mu^\prime)\bigr)d(\mu-\mu^\prime)(x)\geq 0. \nn
\ee
The uniqueness result in probabilistic settings is given by \cite[Theorem 3.29]{Carmona-Delarue-1}.
It says that there is at most one MFG equilibrium if the running as well as terminal cost functions satisfy 
Lasry-Lions monotonicity.

On the other hand, the relevant monotonicity  in the current paper is assumed so that 
the continuation method \cite{Peng-Wu} is applicable. 
See, for example, the set of inequalities $(\ref{mfg-mono-summary})$. It essentially corresponds to \cite[(H2.3)]{Peng-Wu} and is used to make Banach's fixed  point theorem applicable to prove the existence as well as the uniqueness of the solution.
\end{remark}

\section{Convergence to the mean-field limit}
From Theorem~\ref{th-mfg-full-existence}, the market-clearing price
in the mean-field limit is naturally expected to be 
\be
\vp_t^{\rm mfg}:=-\mbb{E}_t^0[y^1_t]+\Lambda_t \ol{\calv_t}^0\bigl(-p^0_t+\mbb{E}^0_t[y^1_t]+\mbb{E}^0_t[p^1_t]\bigr), \quad t\in[0,T). 
\label{eq-price-mfg}
\ee
In this section, we shall show that this is indeed the case for the homogeneous minor agents.
Lastly, we also provide the estimate on the difference of the equilibrium price
between the two markets; one is the homogeneous mean-field limit 
and the other is the heterogeneous market of finite population.
\subsection{Large population limit of the minor agents}
We now go back to the original setup of probability space given in Section~\ref{sec-notation}.
We first assume that the minor agents are homogeneous.
\begin{assumption}{\rm (Minor-Homogeneous)}
The conditions in Assumption (MFG) hold true. Moreover, 
every minor agent $1\leq i\leq N$ is subject to the common coefficient functions
$(l, \sigma^0, \sigma, \ol{f}, \ol{g}, c^f, c^g, h^f, h^g)$ given there.
\end{assumption}

For each $1\leq i\leq N$, let us construct $\mbb{F}^i$-adapted processes,
corresponding to those given by $(\ref{eq-mfg-full})$ and $(\ref{eq-mfg-full-terminal})$:
\bea
\begin{cases}
dx_t^0=\Bigl\{\ol{\calv}_t^0(-p^0_t+\mbb{E}^0_t[y^i_t]+\mbb{E}^0_t[p^i_t])+\mg{l}_0(t,c_t^0)\Bigr\}dt+
\mg{s}_0(t,c_t^0)dW_t^0, \\
dx_t^i=\Bigl\{-\ol{\Lambda}_t(y_t^i-\mbb{E}^0_t[y^i_t])-\ol{\calv}_t^0(-p_t^0+\mbb{E}^0_t[y^i_t]+
\mbb{E}^0[p^i_t])+l(t,c_t^0,c_t^i)\Bigr\}dt\\
\qquad +\sigma^0(t,c_t^0,c_t^i)dW_t^0+\sigma(t,c^0_t,c_t^i)dW_t^i, \\
dr_t^i=\Bigl\{\ol{\Lambda}_t(p^i_t-\mbb{E}_t^0[p^i_t])+\ol{\calv}_t^0(-p^0_t+\mbb{E}^0_t[y^i_t]+\mbb{E}^0_t[p^i_t])\Bigr\}dt, \\
dp_t^0=-\part_x \ol{\mg{f}}_0(t,x_t^0,c_t^0)dt+q_t^{0,0}dW_t^0,  \\
dy_t^i=-\part_x \ol{f}(t,x_t^i, c_t^0, c_t^i)dt+z_t^{i,0}dW_t^0+z_t^{i,i}dW_t^i, \\
dp_t^i=c^f(t,c_t^0,c_t^i)r^i_tdt+q_t^{i,0}dW_t^0+q_t^{i,i}dW_t^i, 
\end{cases}
\label{eq-mfg-N-copy}
\eea
with
\bea
\begin{cases}
x_0^0=\chi^0, \quad x^i_0=\xi^i, \quad r_0^i=0, \\
p^0_T=\part_x \mg{g}_0(x_T^0, c_T^0), \\
\display y^i_T=\frac{\del}{1-\del}\mbb{E}^0_T\bigl[c^g(c_T^0,c_T^i)x_T^i+h^g(c_T^0,c_T^i)\bigr]+c^g(c_T^0,c_T^i)x_T^i+h^g(c_T^0,c_T^i), \\
p^i_T=-c^g(c_T^0,c_T^i)\Bigl(r^i_T+\frac{\del}{1-\del}\mbb{E}^0_T[r^i_T]\Bigr).
\end{cases}
\label{eq-mfg-N-copy-terminal}
\eea

By  construction of the probability space and the fact that
 $(c^i, \xi^i, W^i)$ are independently and identically distributed (i.i.d.),  
$\mbb{F}^i$-adapted processes $(x^i, r^i, y^i, p^i)$ are $\ol{\calf}^0$-conditionally i.i.d.
In particular, for any $\varphi^i=x^i, r^i, y^i, p^i$, we have 
$\mbb{E}_t^0[\varphi^i_t]=\mbb{E}^0_t[\varphi^1_t]$
and also
\be
\mbb{E}^0_T\bigl[c^g(c^0_T, c_T^i)x^i_T+h^g(c_T^0+c_T^i)\bigr]=\mbb{E}^0_T\bigl[c^g(c^0_T, c_T^1)x^1_T+h^g(c_T^0+c_T^1)\bigr]. \nn
\ee
Therefore, $(x^0, p^0)$ defined in $(\ref{eq-mfg-N-copy})$ and $(\ref{eq-mfg-N-copy-terminal})$
is indeed indistinguishable for every copy.

We are going to compare $\bigl(x^0, p^0, (x^i)_{i=1}^N , (r^i)_{i=1}^N, (y^i)_{i=1}^N, (p^i)_{i=1}^N \bigr)$ given above with 
the dynamics $\bigl(X^0/N, P^0, (X^i)_{i=1}^N, (R^i)_{i=1}^N, (Y^i)_{i=1}^N, (P^i)_{i=1}^N\bigr)$
given by $(\ref{eq-full-coupled})$ and $(\ref{eq-full-terminal})$
with homogeneous coefficients.
Using the scaling rule in $(\ref{def-scaling})$ and Remark~\ref{remark-derivative}, we have for $1\leq i\leq N$,
\bea
\begin{cases}
\display d \frac{X_t^0}{N}=\Bigl\{\ol{\calv}_t^0\bigl(-P_t^0+\mg{m}((Y_t))+\mg{m}((P_t))\bigr)
+\mg{l}_0(t,c_t^0)\Bigr\}dt+\mg{s}_0(t,c_t^0)dW_t^0, \\
dX_t^i=\Bigl\{-\ol{\Lambda}_t\bigl(Y_t^i-\mg{m}((Y_t))\bigr)-\ol{\calv}_t^0\bigl(-P_t^0+\mg{m}((Y_t))+\mg{m}((P_t))\bigr)
+l(t,c_t^0,c_t^i)\Bigr\}dt \\
\qquad+\sigma^0(t,c_t^0,c_t^i)dW_t^0+\sigma(t,c_t^0,c_t^i)dW_t^i, \\
dR_t^i=\Bigl\{\ol{\Lambda}_t\bigl(P_t^i-\mg{m}((P_t))\bigr)+\ol{\calv}_t^0\bigl(-P_t^0+\mg{m}((Y_t))+\mg{m}((P_t))\bigr)\Bigr\}dt, \\
dP_t^0=-\part_x \ol{\mg{f}}_0(t,X_t^0/N, c_t^0)dt+Q_t^{0,0}dW_t^0+\sum_{j=1}^N Q_t^{0,j}dW_t^j, \\
dY_t^i=-\part_x \ol{f}(t,X_t^i, c_t^0,c_t^i)dt+Z_t^{i,0}dW_t^0+\sum_{j=1}^N Z_t^{i,j}dW_t^j, \\
dP_t^i=c^f(t,c_t^0,c_t^i)R_t^idt+Q_t^{i,0}dW_t^0+\sum_{j=1}^N Q_t^{i,j}dW_t^j, 
\end{cases}
\label{eq-N-homo}
\eea
with
\bea
\begin{cases}
X_0^0=N\chi^0, \quad X^i_0=\xi^i, \quad R^i_0=0, \\
P_T^0=\part_x \mg{g}_0(X_T^0/N, c_T^0), \\
\display Y_T^i=\frac{\del}{1-\del}\mg{m}\Bigl(\bigl(c^g(c_T^0,c_T^j)X_T^j+h^g(c_T^0,c_T^j)\bigr)_{j=1}^N\Bigr)+
c^g(c_T^0,c_T^i)X_T^i+h^g(c_T^0,c_T^i), \\
P_T^i=-c^g(c_T^0,c_T^i)\Bigl(R_T^i+\frac{\del}{1-\del}\mg{m}((R_T))\Bigr).
\end{cases}
\label{eq-N-homo-terminal}
\eea
Thanks to the symmetry, $(X^i, R^i, Y^i, P^i)$ have the same distribution for every $1\leq i\leq N$, 
although they are not independent due to their interactions.
Let us  introduce the notation:
\be
\begin{split}
&\Del x^0_t:=\frac{X_t^0}{N}-x_t^0, \quad \Del x^i_t:=X^i_t-x^i_t, \quad \Del r^i_t:=R^i_t-r^i_t, \\
&\Del p^0_t:=P^0_t-p^0_t, \quad \Del y^i_t:=Y_t^i-y^i_t, \quad \Del p^i_t:=P^i_t-p^i_t, \\
&\Del q_t^{0,0}:=Q_t^{0,0}-q_t^{0,0}, \quad \Del q_t^{0,j}:=Q_t^{0,j}, \\
&\Del z_t^{i,0}:=Z_t^{i,0}-z_t^{i,0}, \quad \Del z_t^{i,j}:=Z_t^{i,j}-\del_{i,j}z_t^{i,i}, \\
&\Del q_t^{i,0}:=Q_t^{i,0}-q_t^{i,0}, \quad \Del q_t^{i,j}:=Q_t^{i,j}-\del_{i,j}q_t^{i,i},
\end{split}\nn
\ee
where $\del_{i,j}$ stands for  Kronecker delta.
We also define
\be
\begin{split}
&\ol{\mu}_t^{r,N}:=\frac{1}{N}\sum_{i=1}^N \del_{r^i_t}, \quad \ol{\mu}_t^{y,N}:=\frac{1}{N}\sum_{i=1}^N \del_{y^i_t},  
\quad \ol{\mu}_t^{p, N}:=\frac{1}{N}\sum_{i=1}^N \del_{p^i_t},
\quad \ol{\mu}^{g,N}:=\frac{1}{N}\sum_{i=1}^N \del_{c^g(c_T^0,c_T^i)x_T^i+h^g(c_T^0,c_T^i)},  \\
&\mu^r_t:=\call(r^1_t|\ol{\calf}_t^0),\quad \mu^y_t:=\call(y^1_t|\ol{\calf}_t^0), 
 \quad \mu_t^p:=\call(p^1_t|\ol{\calf}_t^0), 
\quad \mu^g:=\call(c^g(c_T^0,c_T^1)x_T^1+h^g(c_T^0,c_T^1)|\ol{\calf}_T^0).
\end{split} \nn
\ee
Here, $\ol{\mu}^{r,N}, \ol{\mu}^{y,N}, \ol{\mu}^{p,N}$ and  $\ol{\mu}^{g,N}$ denote the empirical measures,
and the others conditional distributions. When the filtration defined on the product space 
is completed, there appears some subtle issue on the conditional distribution about its measurability. 
However, we can always construct a measurable version by modifying it only on the null sets.
We always suppose that $(\mu^r, \mu^y, \mu^p, \mu^g)$ are measurable versions constructed in such a way.
See Section 2.1.3 in \cite{Carmona-Delarue-2} for details.
Since $\bigl(r^i, y^i, p^i, c^g(c_T^0,c_T^i)x^i_T+h^g(c_T^0,c_T^i)\bigr), 1\leq i\leq N$
are $\ol{\calf}^0$ conditionally i.i.d. and also $(r^i, y^i, p^i)$ are continuous processes,
we have the following convergence properties.
\begin{lemma}
\label{lemma-Glivenko-Cantelli}
Let Assumption (Minor-Homogeneous) be in force.
Then we have
\be
\begin{split}
&\lim_{N\rightarrow \infty} \sup_{t\in[0,T]}\mbb{E}\Bigl[W_2(\ol{\mu}_t^{r,N}, \mu_t^r)^2+
W_2(\ol{\mu}^{y,N}_t, \mu_t^y)^2+W_2(\ol{\mu}^{p,N}_t, \mu_t^p)^2\Bigr]=0,  \\
&\lim_{N\rightarrow \infty}\mbb{E}\bigl[W_2(\mu^{g,N},\mu^g)^2\bigr]=0.
\end{split} \nn
\ee
Moreover, if there exist some positive constants $\Gamma$ and $\Gamma_g$ such that
$\sup_{t\in[0,T]}\bigl(\mbb{E}[|r^1_t|^k]^\frac{1}{k}+\mbb{E}[|y^1_t|^k]^\frac{1}{k}+\mbb{E}[|p^1_t|^k]^\frac{1}{k}\bigr)\leq \Gamma$
and $\mbb{E}\bigl[|c^g(c_T^0,c_T^1)x^1_T+h^g(c_T^0,c_T^1)|^k\bigr]^\frac{1}{k}\leq \Gamma_g$
for some $k>4$, then there exists some constant $C$ independent of $N$ such that
\be
\begin{split}
&\sup_{t\in[0,T]}
\mbb{E}\Bigl[W_2(\ol{\mu}_t^{r,N}, \mu_t^r)^2+
W_2(\ol{\mu}^{y,N}_t, \mu_t^y)^2+W_2(\ol{\mu}^{p,N}_t, \mu_t^p)^2\Bigr]
\leq C\Gamma^2 \ep_N, \\
&\quad \mbb{E}\bigl[W_2(\ol{\mu}^{g,N},\mu^g)^2\bigr]\leq C\Gamma^2_g \ep_N, 
\end{split} \nn
\ee
with $\ep_N:=N^{-2/\max(n,4)}(1+\log(N)\bold{1}_{N=4})$.
\begin{proof}
See Lemma~4.1 in \cite{Fujii-mfg-convergence} and the proof for Theorem 5.1 in \cite{Fujii-Takahashi}.
More details on the Glivenko-Cantelli convergence in the Wasserstein distance are available 
from Section 5.1 in \cite{Carmona-Delarue-1} and references therein.
\end{proof}
\end{lemma}

The next property of the Wasserstein distance is important for our purpose.
For any $\mu, \nu\in \calp_2(\mbb{R}^n)$, it is easy to check
\be
\Bigl|\int_{\mbb{R}^n} x \mu(dx)-\int_{\mbb{R}^n} y \nu(dy)\Bigr|=\Bigl|\int_{\mbb{R}^{n\times  n}}
(x-y)\pi(dx,dy)\Bigr|\leq \int_{\mbb{R}^{n\times n}}|x-y|\pi(dx,dy) \nn
\ee
for any coupling $\pi\in \Pi_2(\mu,\nu)$ with marginals $\mu$ and $\nu$. Taking infimum over $\pi\in \Pi_2(\mu, \nu)$, we get 
\be
\Bigl|\int_{\mbb{R}^n} x \mu(dx)-\int_{\mbb{R}^n} y \nu(dy)\Bigr|\leq W_1(\mu,\nu)\leq W_2(\mu, \nu).
\label{ineq-W2}
\ee
We are now ready to prove the main result of this section.
\begin{theorem}
\label{th-mfg-convergence}
Let Assumption (Minor-Homogeneous) be in force.
Then, for every $1\leq i\leq N$, there exists an $N$-independent constant $C$  such that
\be
\begin{split}
&\mbb{E}\Bigl[\sup_{t\in[0,T]}\bigl(|\Del x_t^0|^2+|\Del x^i_t|^2+|\Del r^i_t|^2+|\Del p^0_t|^2+
|\Del y^i_t|^2+|\Del p^i_t|^2\bigr)  \\
&\qquad\qquad+\sum_{j=0}^N \int_0^T \bigl(|\Del q_t^{0,j}|^2+|\Del z_t^{i,j}|^2+|\Del q_t^{i,j}|^2\bigr)dt\Bigr]\\
&\leq C\mbb{E}\Bigl[W_2(\ol{\mu}^{g,N}, \mu^g)^2+W_2(\ol{\mu}^{r,N}_T,\mu^r_T)^2
+\int_0^T \Bigl(W_2(\ol{\mu}_t^{y,N},\mu_t^y)^2+W_2(\ol{\mu}_t^{p,N},\mu_t^p)^2\Bigr)dt\Bigr].
\end{split}\nn
\ee
\begin{proof}
Let us define $\gamma>0$ by
\be
\gamma:=\min\bigl\{\gamma^f_0, \gamma^f, \gamma^g_0, \gamma^g-\mg{a}\bigr\}. \nn
\ee
{\bf First step}: We want to apply \Ito-formula to
\be
\Bigl(\blangle \Del p^0_t, \Del x^0_t\brangle+\frac{1}{N}\sum_{i=1}^N \blangle \Del y^i_t, \Del x_t^i\brangle
+\frac{1}{N} \sum_{i=1}^{N}\blangle \Del p^i_t, (-I)\Del r_t^i\brangle\Bigr). 
\label{step1-ito}
\ee
With this in mind, we check the following estimates.
It is easy to see, with obvious notation, 
\be
\begin{split}
&\blangle {\rm drift}[\Del p^0_t], \Del x_t^0\brangle+\frac{1}{N}\sum_{i=1}^N
\blangle {\rm drift}[\Del y^i_t], \Del x_t^i\brangle+\frac{1}{N}\sum_{i=1}^N \blangle {\rm drift}[\Del p^i_t],
(-I)\Del r^i_t\brangle \\
&\leq -\gamma^f_0 |\Del x_t^0|^2-\gamma^f \frac{1}{N}\sum_{i=1}^N \bigl(|\Del x_t^i|^2+|\Del r_t^i|^2\bigr)
\end{split}
\label{step1-drift-1}
\ee
Using $(\ref{ineq-W2})$, we get
\be
\begin{split}
\blangle {\rm drift}[\Del x_t^0], \Del p_t^0\brangle
&=\blangle \ol{\calv}_t^0\bigl(-\Del p_t^0+\mg{m}((Y_t))-\mbb{E}_t^0[y_t^1]+\mg{m}((P_t))-\mbb{E}_t^0[p_t^1]\bigr), \Del p_t^0\brangle \\
&=\blangle \ol{\calv}_t^0\bigl(-\Del p_t^0+\mg{m}((\Del y_t))+\mg{m}((\Del p_t))\bigr),\Del p_t^0\brangle \\
&\qquad +\blangle \ol{\calv}_t^0\bigl(\mg{m}((y_t))-\mbb{E}_t^0[y_t^1]+\mg{m}((p_t))-\mbb{E}_t^0[p_t^1]\bigr), \Del p^0_t\brangle \nn \\
&\leq \blangle \ol{\calv}_t^0\bigl(-\Del p_t^0+\mg{m}((\Del y_t))+\mg{m}((\Del p_t))\bigr),\Del p_t^0\brangle \\
&\qquad +C\bigl(W_2(\ol{\mu}_t^{y,N},\mu_t^y)+W_2(\ol{\mu}_t^{p,N}, \mu_t^p)\bigr)|\Del p^0_t|. 
\end{split}\nn
\ee
Similar calculation yields
\be
\begin{split}
&\blangle {\rm drift}[\Del x_t^0], \Del p_t^0\brangle
+\frac{1}{N}\sum_{i=1}^N \blangle {\rm drift}[\Del x_t^i], \Del y_t^i\brangle+
\frac{1}{N}\sum_{i=1}^N \blangle {\rm drift}[\Del r_t^i], (-I)\Del p_t^i\brangle\\
&\quad \leq -\blangle \ol{\calv}_t^0\bigl(-\Del p^0_t+\mg{m}((\Del y_t))+\mg{m}((\Del p_t))\bigr), -\Del p^0_t+\mg{m}((\Del y_t))
+\mg{m}((\Del p_t))\brangle\\
&\qquad+C\bigl(W_2(\ol{\mu}_t^{y,N},\mu_t^y)+W_2(\ol{\mu}_t^{p,N},\mu_t^p)\bigr)\bigl(|\Del p^0_t|+|\mg{m}((\Del y_t))|
+|\mg{m}((\Del p_t))|\bigr) \\
&\quad \leq C\bigl(W_2(\ol{\mu}_t^{y,N},\mu_t^y)+W_2(\ol{\mu}_t^{p,N},\mu_t^p)\bigr)\bigl(|\Del p^0_t|+\mg{m}((|\Del y_t|))|
+\mg{m}((|\Del p_t|))\bigr).
\end{split} 
\label{step1-drift-2}
\ee
Now, let us check the terminal parts. Similar analysis used in  $(\ref{eq-terminal-cal})$ 
yields
\be
\begin{split}
&\blangle \Del p^0_T, \Del x_T^0\brangle+\frac{1}{N}\sum_{i=1}^N \blangle \Del y^i_T, \Del x_T^i\brangle
+\frac{1}{N}\sum_{i=1}^N \blangle \Del p^i_T, (-I)\Del r^i_T\brangle \\
&\geq \gamma^g_0|\Del x_T^0|^2+(\gamma^g-\mg{a})\frac{1}{N}\sum_{i=1}^N \bigl(|\Del x_T^i|^2+|\Del r^i_T|^2\bigr)\\
&\quad-CW_2(\ol{\mu}^{g,N},\mu^g)\mg{m}((|\Del x_T|))-CW_2(\ol{\mu}^{r,N}_T, \mu^r_T)\mg{m}((|\Del r_T|)).
\end{split}
\label{step1-terminal}
\ee

From $(\ref{step1-drift-1})$ and $(\ref{step1-drift-2})$, we get
\be
\begin{split}
&\mbb{E}\Bigl[\int_0^T d\Bigl(\blangle \Del p^0_t, \Del x^0_t\brangle+\frac{1}{N}\sum_{i=1}^N \blangle \Del y^i_t, \Del x_t^i\brangle
+\frac{1}{N} \sum_{i=1}^{N}\blangle \Del p^i_t, (-I)\Del r_t^i\brangle\Bigr)\Bigr]\\
&\leq -\gamma \mbb{E}\Bigl[\int_0^T  \Bigl(|\Del x_t^0|^2+\frac{1}{N}\sum_{i=1}^N \bigl(|\Del x_t^i|^2+|\Del r_t^i|^2\bigr)\Bigr)dt\Bigr]  \\
&\quad +C\mbb{E}\Bigl[\int_0^T \bigl(W_2(\ol{\mu}_t^{y,N},\mu_t^y)+W_2(\ol{\mu}^{p,N}_t,\mu^p_t)\bigr)
\bigl(|\Del p^0_t|+\mg{m}((|\Del y_t|))+\mg{m}((|\Del p_t|))\bigr)dt\Bigr].
\end{split} \nn
\ee
Note that there is no quadratic covariation term.
Now combining the estimate $(\ref{step1-terminal})$, we obtain the following:
\be
\begin{split}
&\gamma \mbb{E}\Bigl[|\Del x_T^0|^2+\frac{1}{N}\sum_{i=1}^N \bigl(|\Del x_T^i|^2+|\Del r_T^i|^2\bigr)
+\int_0^T \Bigl(|\Del x_t^0|^2+\frac{1}{N}\sum_{i=1}^N \bigl(|\Del x_t^i|^2+|\Del r_t^i|^2\bigr)\Bigr)dt\Bigr] \\
&\leq C\mbb{E}\Bigl[W_2(\ol{\mu}^{g,N},\mu^g)\mg{m}((|\Del x_T|))+W_2(\ol{\mu}^{r,N}_T,\mu^r_T)\mg{m}((|\Del r_T|))\\
&\qquad\quad+\int_0^T \bigl(W_2(\ol{\mu}_t^{y,N},\mu_t^y)+W_2(\ol{\mu}^{p,N}_t,\mu^p_t)\bigr)
\bigl(|\Del p^0_t|+\mg{m}((|\Del y_t|))+\mg{m}((|\Del p_t|))\bigr)dt\Bigr]. 
\end{split} \nn
\ee
By Young's inequality and the symmetry of the distribution, 
we find that the following inequality holds for every $1\leq i\leq N$:
\be
\begin{split}
&\mbb{E}\Bigl[|\Del x_T^0|^2+|\Del x_T^i|^2+|\Del r_T^i|^2
+\int_0^T \bigl(|\Del x_t^0|^2+|\Del x_t^i|^2+|\Del r_t^i|^2\bigr)dt\Bigr] \\
&\leq C\mbb{E}\Bigl[W_2(\ol{\mu}^{g,N},\mu^g)^2+W_2(\ol{\mu}^{r,N}_T,\mu^r_T)^2 \\
&\qquad\quad+\int_0^T \bigl(W_2(\ol{\mu}_t^{y,N},\mu_t^y)+W_2(\ol{\mu}^{p,N}_t,\mu^p_t)\bigr)
\bigl(|\Del p^0_t|+\mg{m}((|\Del y_t|))+\mg{m}((|\Del p_t|))\bigr)dt\Bigr]. 
\label{step1-final-conv}
\end{split}
\ee
\\
{\bf Second step}\\
From the standard estimate of the BSDEs, see Section 4.4 in \cite{Zhang-BSDE} for example, 
it is easy to find
\be
\begin{split}
&\mbb{E}\Bigl[\sup_{t\in[0,T]}|\Del p_t^0|^2+\int_0^T \Bigl(|\Del q_t^{0,0}|^2+\sum_{j=1}^N |\Del q_t^{0,j}|^2\Bigr)dt\Bigr]\\
&\quad\leq C\mbb{E}\Bigl[ |\Del p_T^0|^2+\int_0^T |\part_x \ol{\mg{f}}_0(t,X_t^0/N, c_t^0)
-\ol{\mg{f}}_0(t,x_t^0,c_t^0)|^2 dt\Bigr]\\
&\quad \leq C\mbb{E}\Bigl[|\Del x_T^0|^2+\int_0^T |\Del x_t^0|^2 dt\Bigr]. 
\end{split}\nn
\ee
Carrying out the similar analysis for $(\Del y^i, \Del p^i)$ and 
using the symmetry among $1\leq i\leq N$, we obtain for any $1\leq i\leq N$,
\be
\begin{split}
&\mbb{E}\Bigl[\sup_{t\in[0,T]}\bigl( |\Del p^0_t|^2+|\Del y_t^i|^2+|\Del p_t^i|^2\bigr)
+\sum_{j=0}^N \int_0^T \bigl(|\Del q_t^{0,j}|^2+|\Del z_t^{i,j}|^2+|\Del q_t^{i,j}|^2\bigr)dt\Bigr]\\
& \leq C\mbb{E}\Bigl[|\Del x_T^0|^2+|\Del x_T^i|^2+|\Del r_T^i|^2+\int_0^T  \bigl(|\Del x_t^0|^2+|\Del x_t^i|^2+|\Del r_t^i|^2\bigr)dt
\Bigr]\\
&\qquad\quad +C\mbb{E}\Bigl[W_2(\ol{\mu}^{g,N},\mu^g)^2+W_2(\ol{\mu}^{r,N}_T, \mu^r_T)^2\Bigr]\\
&\leq C\mbb{E}\Bigl[W_2(\ol{\mu}^{g,N},\mu^g)^2+W_2(\ol{\mu}^{r,N}_T,\mu^r_T)^2 \\
&\qquad\quad+\int_0^T \bigl(W_2(\ol{\mu}_t^{y,N},\mu_t^y)+W_2(\ol{\mu}^{p,N}_t,\mu^p_t)\bigr)
\bigl(|\Del p^0_t|+\mg{m}((|\Del y_t|))+\mg{m}((|\Del p_t|))\bigr)dt\Bigr],
\label{step2-middle-conv}
\end{split}\nn
\ee
where we have used $(\ref{step1-final-conv})$ in the second inequality.

From $(\ref{step1-final-conv})$, $(\ref{step2-middle-conv})$ and the symmetry among $1\leq i\leq N$, 
Young's inequality yields
\be
\begin{split}
&\mbb{E}\Bigl[|\Del x_T^0|^2+|\Del x_T^i|^2+|\Del r_T^i|^2
+\int_0^T \bigl(|\Del x_t^0|^2+|\Del x_t^i|^2+|\Del r_t^i|^2\bigr)dt\Bigr] \\
&\quad +\mbb{E}\Bigl[\sup_{t\in[0,T]}\bigl( |\Del p^0_t|^2+|\Del y_t^i|^2+|\Del p_t^i|^2\bigr)
+\sum_{j=0}^N \int_0^T \bigl(|\Del q_t^{0,j}|^2+|\Del z_t^{i,j}|^2+|\Del q_t^{i,j}|^2\bigr)dt\Bigr]\\
&\leq C\mbb{E}\Bigl[W_2(\ol{\mu}^{g,N}, \mu^g)^2+W_2(\ol{\mu}^{r,N}_T,\mu^r_T)^2
+\int_0^T \Bigl(W_2(\ol{\mu}_t^{y,N},\mu_t^y)^2+W_2(\ol{\mu}_t^{p,N},\mu_t^p)^2\Bigr)dt\Bigr].
\end{split} \nn
\ee
Now the desired estimate follows from a simple application of Burkholder-Davis-Gundy (BDG)
inequality to the forward variables $(\Del x^0, \Del x^i, \Del r^i)$.
\end{proof}
\end{theorem}

\subsection{Some stability results}

For understanding the implications of  Lemma~\ref{lemma-Glivenko-Cantelli} and Theorem~\ref{th-mfg-convergence},
let us denote the market-clearing price for the $N$ homogeneous minor agents by
\bea
\vp_t^{{\rm Ho}, N}:=-\mg{m}((Y_t))+\ol{\calv}_t^0\bigl(-P_t^0+\mg{m}((Y_t))+\mg{m}((P_t))\bigr), \quad t\in[0,T)
\eea
using the solution to $(\ref{eq-N-homo})$ with $(\ref{eq-N-homo-terminal})$.
By comparing it with $\vp_t^{\rm mfg}$ in $(\ref{eq-price-mfg})$, we get the following result.
\begin{theorem}
\label{th-mfg-ho}
Under Assumption (Minor-Homogeneous), the following inequality holds:
\be
\begin{split}
&\mbb{E} \int_0^T\bigl|\vp_t^{{\rm Ho}, N}-\vp_t^{\rm mfg}\bigr|^2dt \\
&\quad \leq C\mbb{E}\Bigl[W_2(\ol{\mu}^{g,N}, \mu^g)^2+W_2(\ol{\mu}^{r,N}_T,\mu^r_T)^2
+\int_0^T \Bigl(W_2(\ol{\mu}_t^{y,N},\mu_t^y)^2+W_2(\ol{\mu}_t^{p,N},\mu_t^p)^2\Bigr)dt\Bigr], 
\end{split}\nn
\ee
where $C$ is some positive constant independent of $N$.
\begin{proof}
Using the symmetry,  we have 
\be
\mbb{E}\bigl[|\vp_t^{{\rm Ho}, N}-\vp_t^{\rm mfg}|^2\bigr]\leq C\mbb{E}\Bigl[|\Del p^0_t|^2+|\Del y_t^1|^2+|\Del p^1_t|^2
+W_2(\ol{\mu}^{y,N}_t, \mu_t^y)^2+W_2(\ol{\mu}_t^{p,N}, \mu_t^p)^2\Bigr].\nn
\ee
Hence Theorem~\ref{th-mfg-convergence} gives the desired estimate.
\end{proof}
\end{theorem}

From Lemma~\ref{lemma-Glivenko-Cantelli}, we observe that $(\vp^{{\rm Ho},N}_t)_{t\in[0,T]}$
converges to $(\vp^{\rm mfg}_t)_{t\in[0,T]}$
in the large population limit of homogeneous minor agents.
In this limit, the optimization problem for each $i$th minor agent given in $(\ref{eq-adjoint-minor})$
is solved within $(\Omega^i, \calf^i, \mbb{P}^i; \mbb{F}^i)$
since the market price process $\vp^{\rm mfg}$ is now $\ol{\mbb{F}}^0$-adapted i.e.
dependent only on the common market information. One can observe that 
the natural information structure mentioned in Remark~\ref{remark-information-structure}
is actually achieved in the mean-field limit.

Before closing the paper, let us briefly discuss about the stability relation 
between the heterogeneous and the homogeneous market.
Let $(X^0, (X^i)_{i=1}^N, (R^i)_{i=1}^N, P^0, (P^i)_{i=1}^N)$ denote the unique solution 
to $(\ref{eq-full-coupled})$ with $(\ref{eq-full-terminal})$ given by
Theorem~\ref{th-N-major-existence} in the market with heterogeneous minor agents,
and $(\ul{X}^0, (\ul{X}^i)_{i=1}^N, (\ul{R}^i)_{i=1}^N, \ul{P}^0, (\ul{P}^i)_{i=1}^N)$ the unique solution 
to $(\ref{eq-N-homo})$ with $(\ref{eq-N-homo-terminal})$ corresponding to the homogeneous minor agents.
Let us introduce the following notation: for $1\leq i\leq N$, 
\be
\begin{split}
&\del l_i(t):=l_i(t,c_t^0,c_t^i)-l(t,c_t^0, c_t^i), \\
&\del \sigma_i^0(t):=\sigma_i^0(t,c_t^0,c_t^i)-\sigma^0(t,c_t^0,c_t^i), 
\quad \del \sigma_i(t):=\sigma_i(t,c_t^0,c_t^i)-\sigma(t,c_t^0,c_t^i), \\
&\del \part_x \ol{f}_i(t):=\part_x \ol{f}_i(t,\ul{X}_t^i,c_t^0,c_t^i)-\part_x \ol{f}(t,\ul{X}_t^i,c_t^0,c_t^i), \\
&\del c_i^f(t):=c_i^f(t,c_t^0,c_t^i)-c^f(t,c_t^0,c_t^i), \\
&\del h_i^g=h_i^g(c_T^0,c_T^i)-h^g(c_T^0,c_T^i).
\end{split}\nn
\ee
Denoting the market-clearing price $(\ref{eq-price-hetero})$ in the market with $N$ heterogeneous agents
by $(\vp_t^{{\rm He}, N})_{t\in[0,T)}$, we have the next stability result.
\begin{corollary}
\label{coro-hetero}
Let Assumptions (Minor-A, B) and (MFG) be in force. 
Then the following inequality holds:
\be
\begin{split}
&\mbb{E}\int_0^T \bigl|\vp_t^{{\rm He}, N}-\vp_t^{\rm mfg}|^2 dt\\
&\quad \leq C\mbb{E}\Bigl[W_2(\ol{\mu}^{g,N}, \mu^g)^2+W_2(\ol{\mu}^{r,N}_T,\mu^r_T)^2
+\int_0^T \Bigl(W_2(\ol{\mu}_t^{y,N},\mu_t^y)^2+W_2(\ol{\mu}_t^{p,N},\mu_t^p)^2\Bigr)dt\Bigr]\\
&\quad+C\frac{1}{N}\sum_{i=1}^N \mbb{E}\int_0^T
\Bigl(|\part_x \ol{f}_i(t)|^2+|\del c^f_i(t)\ul{R}_t^i|^2+|\del l_i(t)|^2+|\del \sigma^0_i(t)|^2+|\del \sigma_i(t)|^2\Bigr)dt\\
&\quad+C\frac{1}{N}\sum_{i=1}^N \mbb{E}\Bigl[
|\del c_i^g \ul{X}_T^i+\del h_i^g|^2+\Bigl|\del c_i^g \Bigl(\ul{R}_T^i+\frac{\del}{1-\del}\mg{m}((\ul{R}_T))\Bigr)\Bigr|^2\Bigr].
\end{split}\nn
\ee
\begin{proof}
Let us put $\Del X_t^0:=X_t^0-\ul{X}_t^0$, $\Del Y_t^i=Y_t^i-\ul{Y}_t^i$, and 
similarly for the others.
Thanks to the  stability of fully-coupled FBSDEs, see for example Proposition~3.1 in \cite{Fujii-mfg-convergence}
or more generally Proposition~3.4 in \cite{J-Yong-FBSDE}, we have
\be
\begin{split}
&\frac{1}{N}\sum_{i=1}^N \mbb{E}\Bigl[\sup_{t\in[0,T]}\Bigl(\Bigl|\frac{\Del X_t^0}{N}\Bigr|^2+|\Del X_t^i|^2
+|\Del R_t^i|^2+|\Del P^0_t|^2+|\Del Y_t^i|^2+|\Del P_t^i|^2\Bigr)\\
&\qquad\quad+\sum_{j=0}^N\int_0^T \bigl(|\Del Q_t^{0,j}|^2+|\Del Z_t^{i,j}|^2+|\Del Q_t^{i,j}|^2\bigr)dt\Bigr]\\
&\leq C\frac{1}{N}\sum_{i=1}^N \mbb{E}\int_0^T \Bigl(|\del \part_x \ol{f}_i(t)|^2+|\del c_i^f(t)\ul{R}_t^i|^2
+|\del l_i(t)|^2+|\del \sigma_i^0(t)|^2+|\del \sigma_i(t)|^2\Bigr)dt\\
&\quad+C\frac{1}{N}\sum_{i=1}^N \mbb{E}\Bigl[
|\del c_i^g \ul{X}_T^i+\del h_i^g|^2+\Bigl|\del c_i^g \Bigl(\ul{R}_T^i+\frac{\del}{1-\del}\mg{m}((\ul{R}_T))\Bigr)\Bigr|^2\Bigr].
\end{split}
\label{eq-stability-hetero}
\ee
Since 
\be
\mbb{E}\bigl|\vp_t^{{\rm He}, N}-\vp_t^{{\rm Ho}, N}\bigr|^2\leq C\mbb{E}\Bigl[
|\Del P_t^0|^2+\frac{1}{N}\sum_{i=1}^N (|\Del Y_t^i|^2+|\Del P_t^i|^2)\Bigr], \nn
\ee
the estimate $(\ref{eq-stability-hetero})$ and Theorem~\ref{th-mfg-ho} give the desired inequality.
\end{proof}
\end{corollary}

\subsection{Mean-field limit as an approximation}
By the MFG theory for the standard Nash-game settings, it is well-known that the equilibrium strategy in the mean-field limit
provides an $\ep_N$-Nash equilibrium for the corresponding finite $N$-agent game~\cite{Carmona-Delarue-1, Carmona-Delarue-2}.
For the market-clearing equilibrium, the results of this section allow us to obtain not only the accuracy of the approximation 
but also the strong convergence in the large $N$-limit. 
In fact, Theorem~\ref{th-mfg-ho}  combined 
with Lemma~\ref{lemma-Glivenko-Cantelli} provides not only the accuracy of $\vp^{\rm mfg}$
as an approximation
but also the convergence speed of the true price process $\vp^{{\rm Ho}, N}$ in the finite (homogeneous) agent market.

Let us also mention about the trading strategy in the equilibrium.
As one can imagine, the equilibrium strategy in the mean field limit
\be
\begin{split}
&\wh{\beta}_t^{\rm mfg}:=\ol{\calv}_t^0\bigl(-p_t^0+\mbb{E}_t^0[y_t^1]+\mbb{E}_t^0[p_t^1]\bigr),\\
&\wh{\alpha}_t^{{\rm mfg},i}:=-\ol{\Lambda}_t\bigl(y_t^i-\mbb{E}_t^0[y_t^i]\bigr)-\ol{\calv}_t^0\bigl(-p_t^0+\mbb{E}_t^0[y_t^i]+\mbb{E}_t^0[p_t^i]\bigr),
~1\leq i\leq N
\end{split}
\ee
gives an approximate  strategy for the finite agent market, where $\ol{\calv}_t^0=(\Lambda^0+2\Lambda_t)^{-1}$
and $(p^0, y^i, p^i)$ is the solution to the McKean-Vlasov FBSDE $(\ref{eq-mfg-N-copy})$.
Note that $\varphi^i=x^i, r^i, y^i, p^i$, we have $\mbb{E}_t^0[\varphi^i_t]=\mbb{E}^0_t[\varphi^1_t]$ for every $1\leq i\leq N$
and more importantly that $\wh{\beta}^{\rm mfg}$ and $\wh{\alpha}^{{\rm mfg}, i}$ are $\ol{\mbb{F}}^0$- and $\mbb{F}^i$-adapted,
respectively. This means that each agent can implement an approximate strategy without knowing the idiosyncratic 
information for the other agents.
From the result of Section~\ref{sec-finite-agent}, the true equilibrium strategy in the $N$-agent (homogeneous) market is given by
\be
\begin{split}
&\wh{\beta}_t^{N}/N:=\ol{\calv}_t^0\bigl(-P_t^0+\mg{m}((Y_t))+\mg{m}((P_t))\bigr), \\
&\wh{\alpha}_t^{N,i}:=-\ol{\Lambda}_t\bigl(Y_t^i-\mg{m}((Y_t))\bigr)-\ol{\calv}_t^0\bigl(-P_t^0+\mg{m}((Y_t))+\mg{m}((P_t))\bigr), ~1\leq i\leq N
\end{split}
\ee
where $(P^0, Y^i, P^i), 1\leq i\leq N$ is the solution to the $N$-coupled system of FBSDEs $(\ref{eq-N-homo})$.
Observe that every agent needs the perfect information $\mbb{F}$ to implement the strategy in this case.
The accuracy of $(\wh{\beta}_t^{\rm mfg},\wh{\alpha}_t^{{\rm mfg},i})$ as an approximation 
for the true strategy  $(\wh{\beta}_t^{N}/N, \wh{\alpha}_t^{N,i})$ can be derived from 
Theorem~\ref{th-mfg-convergence}. In fact, the estimate of the square difference becomes essentially the same as for the equilibrium price process.
In the case of heterogeneous minor agents, one can make use of the stability property of FBSDE as in 
Corollary~\ref{coro-hetero}.

Although it is difficult to obtain a numerical solution for McKean-Vlasov FBSDE $(\ref{eq-mfg-N-copy})$, it looks at least more hopeful
than for the large coupled system of FBSDEs $(\ref{eq-N-homo})$. In fact, the numerical approximation of mean field games
has been one of the hot topics among the researchers in recent years. Moreover, if we adopt an appropriate linear-quadratic cost functions
both for the major and minor agents, we may obtain an explicit form of the solution. Let us leave this problem
as the potential future project.

\section{Securities with maturity $T$}
Let us briefly discuss the special case where the securities have  exogenously 
specified payoff $c_T^0\in \mbb{L}^2(\ol{\calf}_T^0;\mbb{R}^n)$ at the date of maturity $T$.
This is the situation arising in Futures, Bonds and many other financial derivatives.
In this case, there is no reason to put a penalty on the  terminal stock.
It is then natural to consider
\be
\begin{split}
&g_i(x,c^0):=-\blangle c^0,x\brangle, \quad 1\leq i\leq N \\
&g_0^{(N)}(x,c^0)=\mg{g}_0(x,c^0):=-\blangle c^0, x\brangle,
\end{split} \nn
\ee
(i.e. $c_i^g(\cdot)=0$) for the terminal condition for the minor and the major agents, respectively.
Since the terminal costs are linear in $x$, we now have $\gamma_0^g=\gamma^g=0$.
Moreover, we remove the hard constraint  $\beta_T=0$
from the major agent's admissible strategies. It does not play any role since there is no $\vp$ dependence 
in the terminal cost functions for all the players. This means $\mbb{A}^0=\mbb{H}^2(\mbb{F};\mbb{R}^n)$
and $\mbb{A}^0_{\rm mfg}=\mbb{H}^2(\ol{\mbb{F}}^0;\mbb{R}^n)$.

Although we loose strict convexity in the terminal functions, we can actually obtain the same conclusions
also for this case. As we have already mentioned in \cite{Fujii-Takahashi},
what we have to do is to apply Theorem~2.3 instead of Theorem~2.6 in \cite{Peng-Wu}.
Every theorem concerning the existence of the unique solution holds with the 
new terminal condition for the backward variables:
\bea
\begin{cases}
&P_T^0=-c_T^0,\\
&Y_T^i=-c_T^0, \\
&P^i_T=0.
\end{cases}\nn
\eea
for the model with finite number of agents, and
\bea
\begin{cases}
p^0_T=-c_T^0, \\
y_T^1=-c_T^0, \\
p_T^1=0,
\end{cases}\nn
\eea
for the model in the mean-field limit. Note that the verification theorem such as Theorems~\ref{th-mfg-verification}
and \ref{th-A-verification} remain true since they do not require strict convexity in the terminal functions.
In particular, $(\ref{eq-mfg-terminal-convexity})$ holds true with equality.

One can easily check that the market-clearing price satisfies $\vp_T=c_T^0$
in the both cases. 
The estimate in  Theorem~\ref{th-mfg-convergence}
is now given by, for every $1\leq i\leq N$, 
\be
\begin{split}
&\mbb{E}\Bigl[\sup_{t\in[0,T]}\bigl(|\Del x_t^0|^2+|\Del x^i_t|^2+|\Del r^i_t|^2+|\Del p^0_t|^2+
|\Del y^i_t|^2+|\Del p^i_t|^2\bigr)  \\
&\qquad\qquad+\sum_{j=0}^N \int_0^T \bigl(|\Del q_t^{0,j}|^2+|\Del z_t^{i,j}|^2+|\Del q_t^{i,j}|^2\bigr)dt\Bigr]\\
&\leq C\mbb{E}\Bigl[\int_0^T \Bigl(W_2(\ol{\mu}_t^{y,N},\mu_t^y)^2+W_2(\ol{\mu}_t^{p,N},\mu_t^p)^2\Bigr)dt\Bigr].
\end{split}\nn
\ee
One can prove it in the same way by using the new terminal condition; $\Del p^0_T=\Del y^i_T=\Del p_T^i=0$.

\begin{appendix}
\section{Sufficient maximum conditions for controlled-FBSDEs}
\label{sec-controlled-fbsde}
Our optimization problem for the major agent requires the maximum principle for a system of controlled-FBSDEs.
The general issues of controlled-FBSDEs
have been studied, in particular, by Yong~\cite{J-Yong-control, J-Yong-FBSDE},
where the second-order necessary conditions are given for non-convex control domain.
In the current paper, we actually need the sufficient conditions (i.e.~verification theorem) rather than the 
necessary conditions. On the other hand, we only need the convex control domain.
Since we cannot find a useful summary in the existing literature, we provide the relevant theorem
in this appendix. For the readers' convenience, 
we provide the theorem under the setup more general than what is actually needed for our purpose.

We let $(\Omega,\calf,\mbb{P},\mbb{F})$ be a complete filtered probability space
satisfying the usual conditions. It supports a $d$-dimensional Brownian motion $W$
and $\calf_0$ may be non-trivial.
Let the control domain $A\subset \mbb{R}^k$ be closed and convex and the space of admissible controls 
is denoted by $\mbb{A}=\mbb{H}^2(\mbb{F};A)$. 
For a given $T>0$, we introduce the following measurable functions:
\bea
&&b: \Omega\times [0,T]\times \mbb{R}^n \times \mbb{R}^m \times \mbb{R}^{m\times d}\times A\rightarrow \mbb{R}^n, \nn\\
&&\sigma: \Omega\times [0,T]\times \mbb{R}^n \times \mbb{R}^m \times \mbb{R}^{m\times d}\times A\rightarrow \mbb{R}^{n\times d}, \nn \\
&&f:\Omega\times  [0,T]\times \mbb{R}^n \times \mbb{R}^m \times \mbb{R}^{m\times d}\times A\rightarrow \mbb{R}^{m}, \nn \\
&& \gamma: \Omega\times \mbb{R}^m \rightarrow \mbb{R}^n, 
\quad \Phi: \Omega\times \mbb{R}^n\rightarrow \mbb{R}^m, \quad \phi: \Omega\times \mbb{R}^m\rightarrow \mbb{R}^m.\nn
\eea
With these coefficient functions, we consider the following controlled system of FBSDEs:
\bea
\begin{cases}
dx_t=b(t,x_t,y_t,z_t, u_t)dt+\sigma(t,x_t,y_t,z_t, u_t)dW_t,  \\
dy_t=f(t,x_t,y_t,z_t,u_t)dt+z_t dW_t,  \\
x_0=\gamma(y_0)+\xi,   \\
y_T=\Phi(x_T)+\phi(y_0), 
\end{cases}
\label{A-eq-controlled-fbsde}
\eea
where $\xi\in \mbb{L}^2(\calf_0;\mbb{R}^n)$ is given.
See \cite{J-Yong-control, J-Yong-FBSDE} for various 
motivations to include the mixed initial-terminal conditions.

We study an optimization problem, $\inf_{\bg{u}\in \mbb{A}}J(\bg{u})$,  with
\bea
J(\bg{u}):=\mbb{E}\Bigl[\int_0^T F(t,x_t,y_t, z_t, u_t)dt+G(x_T)+g(y_0)\Bigr], \nn
\eea
under the dynamic constraints $(\ref{A-eq-controlled-fbsde})$. Here,
\bea
&&F: \Omega\times [0,T]\times \mbb{R}^n\times \mbb{R}^m\times \mbb{R}^{m\times d}\times A\rightarrow \mbb{R}, \nn \\
&&G: \Omega\times \mbb{R}^n\rightarrow \mbb{R}, \quad g: \Omega \times \mbb{R}^m\rightarrow \mbb{R} \nn
\eea
are measurable functions representing the cost for the agent.
The Hamiltonian $H:\Omega \times [0,T]\times \mbb{R}^n \times \mbb{R}^m\times \mbb{R}^{m\times d}
\times \mbb{R}^n\times \mbb{R}^{n\times d}\times \mbb{R}^m\times A \rightarrow \mbb{R}$ is defined by
\bea
H(t,x,y,z,p,q,r,u)&:=&\bigl\langle p, b(t,x,y,z,u)\bigr\rangle+\bigl\langle q, \sigma(t,x,y,z,u)\bigr\rangle+\bigl\langle
r, f(t,x,y,z,u)\bigr\rangle\nn \\
&&+F(t,x,y,z,u), \nn
\eea
where the brackets in the second term in the right-hand side denote a trace operation.

\begin{assumption}
\label{assumption-A-1}
{\rm (i)} For any $(x,y,z,u)\in \mbb{R}^n\times \mbb{R}^m\times \mbb{R}^{m\times d}\times A$,
$(b,\sigma, f, F)$ are $\mbb{F}$-progressively measurable, $(\gamma, g)$ are $\calf_0$-measurable
and $\Phi, \phi, G$ are $\calf_T$-measurable. \\
{\rm (ii)} For any $(u_t)_{t\in[0,T]}\in \mbb{A}$, there exists a unique strong solution 
$(x_t,y_t,z_t)_{t\in[0,T]} \in \mbb{S}^2(\mbb{F};\mbb{R}^n)\times \mbb{S}^2(\mbb{F};\mbb{R}^m)\times \mbb{H}^2(\mbb{F};\mbb{R}^{m\times d})$
to the controlled FBSDE $(\ref{A-eq-controlled-fbsde})$~\footnote{
For the existence of unique solutions to fully-coupled FBSDEs, see \cite{Peng-Wu, J-Yong-FBSDE}.
In particular, the latter deals with the mixed initial-terminal conditions.}.\\
{\rm (iii)} $(b,\sigma,f, \gamma, \Phi,\phi)$ are one-time continuously differentiable 
in $(x,y,z,u)$ with bounded derivatives. \\
{\rm (iv)} $(F, G, g)$ are one-time continuously differentiable in $(x,y,z,u)$ with uniformly 
Lipschitz continuous derivatives. Moreover, for any given $(x,y,z,u)$, these derivatives are square integrable. \\
{\rm (v)} For any $(u_t)_{t\in[0,T]} \in \mbb{A}$, $J(\bg{u})$ is finite. \\
{\rm (vi)} $(G, g)$ are convex  and  $(\gamma, \Phi,\phi)$ are affine functions in $(x,y)$. 
\end{assumption}

\begin{remark}
For a scalar-valued function $f(x)\in \mbb{R}$, we use the convention $f_x(x)=(\part_{x^i}f(x))_{i=1}^n\in \mbb{R}^n$.
For a vector-valued function $f(x)\in \mbb{R}^m$, we use $f_x(x)\in \mbb{R}^{m\times n}$
with $(f_x(x))_{i,j}=(\part_{x^j}f^i(x))$.
\end{remark}

The adjoint equations are given as follows:
\bea
\begin{cases}
dr_t=-H_y(t,x_t,y_t, z_t, p_t, q_t, r_t, u_t)dt-H_z(t,x_t, y_t, z_t, p_t, q_t, r_t, u_t)dW_t,  \\
dp_t=-H_x(t,x_t, y_t,z_t,p_t,q_t,r_t, u_t)dt+q_t dW_t,  \\
r_0=\mbb{E}\bigl[\phi_y(y_0)^\top r_T |\calf_0\bigr]- \gamma_y(y_0)^\top p_0  -g_y(y_0), \\
p_T=-\Phi_x(x_T)^\top r_T+G_x (x_T). 
\end{cases}
\label{A-eq-adjoint}
\eea

\begin{theorem}
\label{th-A-verification}
Let Assumption~\ref{assumption-A-1} be in force.
Suppose that $(\wh{x}_t,\wh{y}_t, \wh{z}_t)_{t\in[0,T]} \in \mbb{S}^2\times \mbb{S}^2\times \mbb{H}^2$ 
is a unique solution to the FBSDE $(\ref{A-eq-controlled-fbsde})$ with 
some admissible control process $(\wh{u}_t)_{t\in[0,T]}\in \mbb{A}$.
Assume that there exists a solution $(\wh{p}_t, \wh{q}_t, \wh{r}_t)_{t\in[0,T]}\in \mbb{S}^2\times \mbb{H}^2\times \mbb{S}^2$
to $(\ref{A-eq-adjoint})$ with inputs $(\wh{x}_t, \wh{y}_t, \wh{z}_t, \wh{u}_t)_{t\in[0,T]}$,
and that  the map
\bea
\mbb{R}^n\times \mbb{R}^m\times \mbb{R}^{m\times d}\times A\ni (x,y,z,u) \mapsto
H(t,x,y,z,\wh{p}_t, \wh{q}_t, \wh{r}_t, u)\in \mbb{R} \nn 
\eea
is jointly convex in $(x,y,z,u)$ and strictly convex in $u$,  $dt\otimes d\mbb{P}$-a.e.  Moreover,  the equality
\bea
H(t,\wh{x}_t, \wh{y}_t, \wh{z}_t,\wh{p}_t, \wh{q}_t, \wh{r}_t, \wh{u}_t)=\inf_{u\in A}H(t,\wh{x}_t,\wh{y}_t, \wh{z}_t, \wh{p}_t, \wh{q}_t,\wh{r}_t, u) \nn
\eea
holds $dt\otimes d\mbb{P}$-a.e. Then, $(\wh{u}_t)_{t\in[0,T]}$ is a unique optimal solution.
\begin{proof}
Let us denote by $(x_t,y_t, z_t)_{t\in[0,T]}$ the unique solution to $(\ref{A-eq-controlled-fbsde})$
with a given control process $(u_t)_{t\in[0,T]}\in \mbb{A}$.
For notational convenience, let us introduce
\bea
&&\theta_t:=(x_t, y_t, z_t), \quad \wh{\theta}_t:=(\wh{x}_t,\wh{y}_t, \wh{z}_t), \quad \wh{\vr}_t:=(\wh{p}_t, \wh{q}_t, \wh{r}_t), \nn \\
&&\wh{\Theta}_t:=(\wh{x}_t,\wh{y}_t, \wh{z}_t, \wh{p}_t, \wh{q}_t, \wh{r}_t), \quad t\in[0,T]. \nn
\eea
Since $(\gamma, \Phi, \phi)$ are affine, we have
\bea
&&\mbb{E}\Bigl[\bigl\langle G_x (\wh{x}_T), x_T-\wh{x}_T\bigr\rangle+\blangle g_y(\wh{y}_0), y_0-\wh{y}_0\brangle\Bigr]\nn \\
&&=\mbb{E}\Bigl[\bigl\langle \wh{p}_T+\Phi_x(\wh{x}_T)^\top \wh{r}_T, x_T-\wh{x}_T\brangle
+\blangle -\wh{r}_0+\mbb{E}[\phi_y(\wh{y}_0)^\top \wh{r}_T|\calf_0]-\gamma_y(\wh{y}_0)^\top \wh{p}_0, y_0-\wh{y}_0\brangle\Bigr]\nn \\
&&=\mbb{E}\Bigl[\blangle \wh{p}_T, x_T-\wh{x}_T\brangle-\blangle \wh{p}_0, x_0-\wh{x}_0\brangle
+\blangle \wh{r}_T, y_T-\wh{y}_T\brangle-\blangle \wh{r}_0, y_0-\wh{y}_0\brangle\Bigr],\nn
\eea
where we have used the relation, for example, $\Phi_x(\wh{x}_T)(x_T-\wh{x}_T)=\Phi(x_T)-\Phi(\wh{x}_T)$.

Now, \Ito-formula gives
\bea
&&\mbb{E}\Bigl[\blangle \wh{p}_T, x_T-\wh{x}_T\brangle-\blangle \wh{p}_0, x_0-\wh{x}_0\brangle
+\blangle \wh{r}_T, y_T-\wh{y}_T\brangle-\blangle \wh{r}_0, y_0-\wh{y}_0\brangle\Bigr] \nn \\
&&=\mbb{E}\int_0^T \Bigl[\langle \wh{p}_t, b(t,\theta_t, u_t)-b(t,\wh{\theta}_t, \wh{u}_t)\brangle
-\blangle H_x(t,\wh{\Theta}_t, \wh{u}_t), x_t-\wh{x}_t\brangle+\blangle \wh{q}_t, \sigma(t,\theta_t, u_t)-\sigma(t,\wh{\theta}_t, \wh{u}_t)\brangle \nn \\
&&\quad+\blangle \wh{r}_t, f(t,\theta_t u_t)-f(t,\wh{\theta}_t, \wh{u}_t)\brangle-
\blangle H_y(t,\wh{\Theta}_t, \wh{u}_t), y_t-\wh{y}_t\brangle-\blangle H_z(t,\wh{\Theta}_t,\wh{u}_t), z_t-\wh{z}_z\brangle\Bigr]dt. \nn
\eea
It is easy to check that the stochastic integration part becomes a true martingale.
Using the convexity of $G$ and $g$, we have
\bea
&&J(\bg{u})-J(\wh{\bg{u}})\nn \\
&&\geq \mbb{E}\Bigl[\bigl\langle G_x (\wh{x}_T), x_T-\wh{x}_T\bigr\rangle+\blangle g_y(\wh{y}_0), y_0-\wh{y}_0\brangle
+\int_0^T \bigl[F(t,\theta_t,u_t)-F(t,\wh{\theta}_t, \wh{u}_t)\bigr]dt \Bigr] \nn \\
&&=\mbb{E}\int_0^T \Bigl[H(t,\theta_t, \wh{\vr}_t, u_t)-H(t,\wh{\Theta}_t, \wh{u}_t)-
\blangle H_x(t,\wh{\Theta}_t, \wh{u}_t), x_t-\wh{x}_t\brangle \nn \\
&&\qquad\quad-\blangle H_y(t,\wh{\Theta}_t, \wh{u}_t), y_t-\wh{y}_t\brangle-\blangle H_z(t,\wh{\Theta}_t, \wh{u}_t), z_t-\wh{z}_t\brangle\Bigr]dt 
\nn \\
&&\geq \mbb{E}\int_0^T \Bigl[H(t,\theta_t, \wh{\vr}_t, u_t)-H(t,\wh{\Theta}_t, \wh{u}_t)-
\blangle H_x(t,\wh{\Theta}_t, \wh{u}_t), x_t-\wh{x}_t\brangle \nn \\
&&\qquad\quad-\blangle H_y(t,\wh{\Theta}_t, \wh{u}_t), y_t-\wh{y}_t\brangle-\blangle H_z(t,\wh{\Theta}_t, \wh{u}_t), z_t-\wh{z}_t\brangle
-\blangle H_u(t,\wh{\Theta}_t, \wh{u}_t), u_t-\wh{u}_t\brangle \Bigr]dt \nn \\
&&\geq 0, \nn
\eea
where the equality hods if and only if $(u=\wh{u})$ due to the strict convexity.
\end{proof}
\end{theorem}

\end{appendix}



\end{document}